\documentclass[12pt, preprint]{aastex}

\newcommand\arcpt{${{\lower3pt\hbox{$^{\prime\prime}$}}\atop{\raise4pt\hbox{.}}}$}

\slugcomment{to appear in the {\it Astronomical Journal}}

\shorttitle{New Nearby Spectroscopically Confirmed White Dwarfs}
\shortauthors{Subasavage et al.}

\begin{document}

\title{The Solar Neighborhood XIX: \\ Discovery and Characterization
of 33 New Nearby White Dwarf Systems}

\author{John P. Subasavage, Todd J. Henry}

\affil{Georgia State University, Atlanta, GA 30302-4106}

\email{subasavage@chara.gsu.edu, thenry@chara.gsu.edu}

\author{P. Bergeron, P. Dufour}

\affil{D\'{e}partement de Physique, Universit\'{e} de Montr\'{e}al,
C.P. 6128, Succ. Centre-Ville, Montr\'{e}al, Qu\'{e}bec H3C 3J7,
Canada}

\email{bergeron@astro.umontreal.ca, dufourpa@astro.umontreal.ca}

\author{Nigel C. Hambly}

\affil{Scottish Universities Physics Alliance (SUPA), Institute for
Astronomy, University of Edinburgh Royal Observatory, Blackford Hill,
Edinburgh EH9 3HJ, Scotland, UK}

\email{nch@roe.ac.uk}

\author{Thomas D. Beaulieu}

\affil{Georgia State University, Atlanta, GA 30302-4106}

\email{beaulieu@chara.gsu.edu}


\begin{abstract}

We present spectra for 33 previously unclassified white dwarf systems
brighter than $V$ $=$ 17 primarily in the southern hemisphere.  Of
these new systems, 26 are DA, 4 are DC, 2 are DZ, and 1 is DQ.  We
suspect three of these systems are unresolved double degenerates.  We
obtained $VRI$ photometry for these 33 objects as well as for 23 known
white dwarf systems without trigonometric parallaxes, also primarily
in the southern hemisphere.  For the 56 objects, we converted the
photometry values to fluxes and fit them to a spectral energy
distribution using the spectroscopy to determine which model to use
(i.e. pure hydrogen, pure helium, or metal-rich helium), resulting in
estimates of $T_{\rm eff}$ and distance.  Eight of the new and 12
known systems are estimated to be within the NStars and Catalogue of
Nearby Stars (CNS) horizons of 25 pc, constituting a potential 18\%
increase in the nearby white dwarf sample.  Trigonometric parallax
determinations are underway via CTIOPI for these 20 systems.

One of the DCs is cool so that it displays absorption in the near
infrared.  Using the distance determined via trigonometric parallax,
we are able to constrain the model-dependent physical parameters and
find that this object is most likely a mixed H/He atmosphere white
dwarf similar to other cool white dwarfs identified in recent years
with significant absorption in the infrared due to collision-induced
absorptions by molecular hydrogen.

\end{abstract}

\keywords{solar neighborhood --- white dwarfs --- stars: evolution ---
stars: distances --- stars: statistics}

\section{Introduction}

The study of white dwarfs (WDs) provides insight to understanding WD
formation rates, evolution, and space density.  Cool WDs, in
particular, provide limits on the age of the Galactic disk and could
represent some unknown fraction of the Galactic halo dark matter.
Individually, nearby WDs are excellent candidates for astrometric
planetary searches because the astrometric signature is greater than
for an identical WD system more distant.  As a population, a complete
volume limited sample is necessary to provide unbiased statistics;
however, their intrinsic faintness has allowed some to escape
detection.

Of the 18 WDs with trigonometric parallaxes placing them within 10 pc
of the Sun (the RECONS sample), all but one have proper motions
greater than 1$\farcs$0 yr$^{-1}$ (94\%).  By comparison, of the 230
main sequence systems (as of 01 January 2007) in the RECONS sample,
50\% have proper motions greater than 1$\farcs$0 yr$^{-1}$.  We have
begun an effort to reduce this apparent selection bias against
slower-moving WDs to complete the census of nearby WDs.  This effort
includes spectroscopic, photometric, and astrometric initiatives to
characterize newly discovered as well as known WDs without
trigonometric parallaxes.  Utilizing the SuperCOSMOS Sky Survey (SSS)
for plate magnitude and proper motion information coupled with data
from other recently published proper motion surveys (primarily in the
southern hemisphere), we have identified relatively bright WD
candidates via reduced proper motion diagrams.

In this paper, we present spectra for 33 newly discovered WD systems
brighter than $V$ $=$ 17.0.  Once an object is spectroscopically
confirmed to be a WD (in this paper for the first time or elsewhere in
the literature), we obtain CCD photometry to derive $T_{\rm eff}$ and
estimate its distance using a spectral energy distribution (SED) fit
and a model atmosphere analysis.  If an object's distance estimate is
within the NStars \citep{2003fst3.book..111H} and CNS
\citep{1991adc..rept.....G} horizons of 25 pc, it is then added to
CTIOPI (Cerro Tololo Inter-American Observatory Parallax
Investigation) to determine its true distance
(e.g.~\citealt*{2005AJ....129.1954J}, \citealt*{2006AJ....132.2360H}).

\section {Candidate Selection}

We used recent high proper motion (HPM) surveys
\citep{2004A&A...421..763P,2005AJ....129..413S,2005AJ....130.1658S,finch}
in the southern hemisphere for this work because our long-term
astrometric observing program CTIOPI, is based in Chile.  To select
good WD candidates for spectroscopic observations, plate magnitudes
via SSS and 2MASS $JHK_S$ are extracted for HPM objects.  Each
object's ($R_{\rm 59F} - J$) color and reduced proper motion (RPM) are
then plotted.  RPM correlates proper motion with proximity, which is
certainly not always true; however, it is effective at separating WDs
from subdwarfs and main sequence stars.  Figure \ref{rpm} displays an
RPM diagram for the 33 new WDs presented here.  To serve as examples
for the locations of subdwarfs and main sequence stars, recent HPM
discoveries from the SuperCOSMOS-RECONS (SCR) proper motion survey are
also plotted \citep{2005AJ....129..413S,2005AJ....130.1658S}.  The
solid line represents a somewhat arbitrary cutoff separating subdwarfs
and WDs.  Targets are selected from the region below the solid line.
Note there are four stars below this line that are not represented
with asterisks.  Three have recently been spectroscopically confirmed
as WDs (Subasavage et al., in preparation) and one as a subdwarf (SCR
1227$-$4541, denoted by ``sd'') that fell just below the line at
($R_{\rm 59F} - J$) $=$ 1.4 and $H_{R_{\rm 59F}}$ $=$ 19.8
\citep{2005AJ....130.1658S}.

Completeness limits (S/N $>$ 10) for 2MASS are $J$ $=$ 15.8, $H$ $=$
15.1, and $K_S$ $=$ 14.3 for uncontaminated point sources
\citep{2006AJ....131.1163S}.  The use of $J$ provides a more reliable
RPM diagram color for objects more than a magnitude fainter than the
$K_S$ limit, which is particularly important for the WDs (with ($J -
K_S$) $<$ 0.4) discussed here.  Only objects bright enough to have
2MASS magnitudes are included in Figure \ref{rpm}.  Consequently, all
WD candidates are brighter than $V$ $\sim$ 17, and are therefore
likely to be nearby.  Objects that fall in the WD region of the RPM
diagram were cross-referenced with SIMBAD and
\citet{1999ApJS..121....1M}\footnote{The current web based catalog can
  be found at http://heasarc.nasa.gov/W3Browse/all/mcksion.html} to
determine those that were previously classified as WDs.  The remainder
were targeted for spectroscopic confirmation.

The remaining 33 candidates comprise the ``new sample'' whose spectra
are presented in this work, while the ``known sample'' constitutes the
23 previously identified WD systems without trigonometric parallaxes
for which we have complete $VRIJHK_S$ data.

\section {Data and Observations}
\subsection {Astrometry and Nomenclature}

The traditional naming convention for WDs uses the object's epoch 1950
equinox 1950 coordinates.  Coordinates for the new sample were
extracted from 2MASS along with the Julian date of observation.  These
coordinates were adjusted to account for proper motion from the epoch
of 2MASS observation to epoch 2000 (hence epoch 2000 equinox 2000).
The coordinates were then transformed to equinox 1950 coordinates
using the IRAF procedure {\it precess}.  Finally, the coordinates were
again adjusted (opposite the direction of proper motion) to obtain
epoch 1950 equinox 1950 coordinates.

Proper motions were taken from various proper motion surveys in
addition to unpublished values obtained via the SCR proper motion
survey while recovering previously known HPM objects.  Appendix A
contains the proper motions used for coordinate sliding as well as
J2000 coordinates and alternate names.

\subsection {Spectroscopy}

Spectroscopic observations were taken on five separate observing runs
in 2003 October and December, 2004 March and September, and 2006 May
at the Cerro Tololo Inter-American Observatory (CTIO) 1.5m telescope
as part of the SMARTS Consortium.  The Ritchey-Chr\'{e}tien
Spectrograph and Loral 1200$\times$800 CCD detector were used with
grating 09, providing 8.6 \AA~resolution and wavelength coverage from
3500 to 6900 \AA.  Observations consisted of two exposures (typically
20 - 30 minutes each) to permit cosmic ray rejection, followed by a
comparison HeAr lamp exposure to calibrate wavelength for each object.
Bias subtraction, dome/sky flat-fielding, and extraction of spectra
were performed using standard IRAF packages.

A slit width of 2$\arcsec$ was used for the 2003 and 2004 observing
runs.  Some of these data have flux calibration problems because the
slit was not rotated to be aligned along the direction of atmospheric
refraction.  In conjunction with telescope ``jitter'', light was
sometimes lost preferentially at the red end or the blue end for these
data.

A slit width of 6$\arcsec$, used for the 2006 May run, eliminated most
of the flux calibration problems even though the slit was not rotated.
All observations were taken at an airmass of less than 2.0.  Within
our wavelength window, the maximum atmospheric differential
refraction is less than 3$\arcsec$ \citep{1982PASP...94..715F}.  A
test was performed to verify that no resolution was lost by taking
spectra of a F dwarf with sharp absorption lines from slit widths of
2$\arcsec$ to 10$\arcsec$ in 2$\arcsec$ increments.  Indeed, no
resolution was lost.

Spectra for the new DA WDs with $T_{\rm eff}$ $\ge$ 10000 K are
plotted in Figure \ref{dahot} while spectra for the new DA WDs with
$T_{\rm eff}$ $<$ 10000 K are plotted in Figure \ref{dacool}.
Featureless DC spectra are plotted in Figure \ref{dc}.  Spectral plots
as well as model fits for unusual objects are described in $\S$ 4.2.

\subsection {Photometry}

Optical $VRI$ (Johnson $V$, Kron-Cousins $RI$) for the new and known
samples was obtained using the CTIO 0.9 m telescope during several
observing runs from 2003 through 2006 as part of the Small and
Moderate Aperture Research Telescope System (SMARTS) Consortium.  The
2048$\times$2046 Tektronix CCD camera was used with the Tek 2 $VRI$
filter set\footnote{The central wavelengths for $V$, $R$, and $I$ are
5475, 6425, and 8075\AA~respectively.}.  Standard stars from
\citet{1982PASP...94..244G}, \citet{1990A&AS...83..357B}, and
\citet{1992AJ....104..340L} were observed each night through a range
of airmasses to calibrate fluxes to the Johnson-Kron-Cousins system and to
calculate extinction corrections.

Bias subtraction and dome flat-fielding (using calibration frames
taken at the beginning of each night) were performed using standard
IRAF packages.  When possible, an aperture 14$\arcsec$ in diameter was
used to determine the stellar flux, which is consistent with the
aperture used by \citet{1992AJ....104..340L} for the standard stars.
If cosmic rays fell within this aperture, they were removed before
flux extraction.  In cases of crowded fields, aperture corrections
were applied and ranged from 4$\arcsec$ to 12$\arcsec$ in diameter
using the largest aperture possible without including contamination
from neighboring sources.  Uncertainties in the optical photometry
were derived by estimating the internal night-to-night variations as
well as the external errors (i.e. fits to the standard stars).  A
complete discussion of the error analysis can be found in
\citet{2004AJ....128.2460H}.  We adopt a total error of $\pm$0.03 mag
in each band.  The final optical magnitudes are listed in Table
\ref{photometry} as well as the number of nights each object was
observed.

Infrared $JHK_S$ magnitudes and errors were extracted via Aladin from
2MASS and are also listed in Table \ref{photometry}.  $JHK_S$
magnitude errors are, in most cases, significantly larger than for
$VRI$, and the errors listed give a measure of the total photometric
uncertainty (i.e. include both global and systematic components).  In
cases when the magnitude error is null, the star is near the magnitude
limit of 2MASS and the photometry is not reliable.

\section {Analysis}
\subsection {Modeling of Physical Parameters}

The pure hydrogen, pure helium, and mixed hydrogen and helium model
atmospheres used to model the WDs are described at length in
\citet{2001ApJS..133..413B} and references therein, while the
helium-rich models appropriate for DQ and DZ stars are described in
\citet{dufour05,dufour07}, respectively. The atmospheric parameters
for each star are obtained by converting the optical $VRI$ and
infrared $JHK_S$ magnitudes into observed fluxes, and by comparing the
resulting SEDs with those predicted from our model atmosphere
calculations. The first step is accomplished by transforming the
magnitudes into average stellar fluxes $f_{\lambda}^m$ received at
Earth using the calibration of \citet{holberg06} for photon counting
devices. The observed and model fluxes, which depend on $T_{\rm eff}$,
$\log g$, and atmospheric composition, are related by the equation

$$f_{\lambda}^m= 4\pi~(R/D)^2~H_{\lambda}^m\ ,\eqno(1)$$

\noindent
where $R/D$ is the ratio of the radius of the star to its distance
from Earth, and $H_{\lambda}^m$ is the Eddington flux, properly
averaged over the corresponding filter bandpass. Our fitting technique
relies on the nonlinear least-squares method of Levenberg-Marquardt
\citep{pressetal92}, which is based on a steepest descent method.  The
value of $\chi ^2$ is taken as the sum over all bandpasses of the
difference between both sides of eq.~(1), weighted by the
corresponding photometric uncertainties.  We consider only $T_{\rm
eff}$ and the solid angle to be free parameters, and the uncertainties
of both parameters are obtained directly from the covariance matrix of
the fit. In this study, we simply assume a value of $\log g=8.0$ for
each star.

As discussed in \citet{BRL, 2001ApJS..133..413B}, the main atmospheric
constituent --- hydrogen or helium --- is determined by comparing the
fits obtained with both compositions, or by the presence of H$\alpha$
in the optical spectra. For DQ and DZ stars, we rely on the procedure
outlined in \citet{dufour05,dufour07}, respectively: we obtain a first
estimate of the atmospheric parameters by fitting the energy
distribution with an assumed value of the metal abundances.  We then
fit the optical spectrum to measure the metal abundances, and 
use these values to improve our atmospheric parameters from
the energy distribution. This procedure is iterated until a
self-consistent photometric and spectroscopic solution is achieved.

The derived values for $T_{\rm eff}$ for each object are listed in
Table \ref{photometry}.  Also listed are the spectral types for each
object determined based on their spectral features.  The DAs have been
assigned a half-integer temperature index as defined by
\citet{1999ApJS..121....1M}, where the temperature index equals
50,400/$T_{\rm eff}$.  As an external check, we compare in Figure
\ref{teffcomp} the photometric effective temperatures for the DA stars
in Table 1 with those obtained by fitting the observed Balmer line
profiles (Figs. \ref{dahot} and \ref{dacool}) using the spectroscopic
technique developed by \citet{1992ApJ...394..228B}, and recently
improved by \citet{liebert03}.  Our grid of pure hydrogen, NLTE, and
convective model atmospheres is also described in Liebert et al.  The
uncertainties of the spectroscopic technique are typically of 0.038
dex in $\log g$ and 1.2\% in $T_{\rm eff}$ according to that study. We
adopt a slightly larger uncertainty of 1.5\% in $T_{\rm eff}$ (Spec)
because of the problematic flux calibrations of the pre$-$2006 data
(see $\S$ 3.2).  The agreement shown in Figure \ref{teffcomp} is
excellent, except perhaps at high temperatures where the photometric
determinations become more uncertain.  It is possible that the
significantly elevated point in Figure \ref{teffcomp}, WD 0310$-$624
(labeled), is an unresolved double degenerate (see $\S$ 4.2).  We
refrain here from using the $\log g$ determinations in our analysis
because these are available only for the DA stars in our sample, and
also because the spectra are not flux calibrated accurately enough for
that purpose.

Once the effective temperature and the atmospheric composition are
determined, we calculate the absolute visual magnitude of each star by
combining the new calibration of \citet{holberg06} with evolutionary
models similar to those described in \citet{fon01} but with C/O cores,
$q({\rm He})\equiv \log M_{\rm He}/M_{\star}=10^{-2}$ and $q({\rm
H})=10^{-4}$ (representative of hydrogen-atmosphere WDs), and $q({\rm
He})=10^{-2}$ and $q({\rm H})=10^{-10}$ (representative of
helium-atmosphere WDs)\footnote{see
http://www.astro.umontreal.ca/\~{}bergeron/CoolingModels/}. By
combining the absolute visual magnitude with the Johnson $V$
magnitude, we derive a first estimate of the distance of each star
(reported in Table \ref{photometry}).  Errors on the distance
estimates incorporate the errors of the photometry values as well as
an error of 0.25 dex in log {\it g}, which is the measured dispersion
of the observed distribution using spectroscopic determinations
\citep[see Figure 9 of][]{1992ApJ...394..228B}.

Of the 33 new systems presented here, 5 have distance estimates within
25 pc.  Four more systems require additional attention because
distance estimates are derived via other means.  Three of these are
likely within 25 pc.  All four are further discussed in the next
section.  In total, 20 WD systems (8 new and 12 known) are estimated
(or determined) to be within 25 pc and one additional common proper
motion binary system possibly lies within 25 pc.

\subsection {Comments on Individual Systems}

Here we address unusual and interesting objects.

{\bf WD 0121$-$429} is a DA WD that exhibits Zeeman splitting of
H$\alpha$ and H$\beta$, thereby making its formal classification DAH.
The SED fit to the photometry is superb, yielding a $T_{\rm eff}$ of
6,369 $\pm$ 137 K.  When we compare the strength of the absorption
line trio with that predicted using the $T_{\rm eff}$ from the SED
fit, the depth of the absorption appears too shallow.  Using the
magnetic line fitting procedure outlined in
\citet{1992ApJ...400..315B}, we must include a 50\% dilution factor to
match the observed central line of H$\alpha$.  In light of this, we
utilized the trigonometric parallax distance determined via CTIOPI of
17.7 $\pm$ 0.7 pc (Subasavage et al., in preparation) to further
constrain this system.  The resulting SED fit, with distance (hence
luminosity) as a constraint rather than a variable, implies a mass of
0.43 $\pm$ 0.03 M$_\odot$.  Given the age of our Galaxy, the lowest
mass WD that could have formed is $\sim$0.47 M$_\odot$
\citep{1984PhR...105..329I}.  It is extremely unlikely that this WD
formed through single star evolution.  The most likely scenario is
that this is a double degenerate binary with a magnetic DA component
and a featureless DC component (necessary to dilute the absorption at
H$\alpha$), similar to G62-46 \citep{1993ApJ...407..733B} and LHS 2273
\citep[see Figure 33 of][]{BRL}.  If this interpretation is correct,
any number of component masses and luminosities can reproduce the SED
fit.

The spectrum and corresponding magnetic fit to the H$\alpha$ lines
(including the dilution) is shown in Figure \ref{lhs1243}.  The
viewing angle, {\it i} $=$ 65$^\circ$, is defined as the angle between
the dipole axis and the line of sight ({\it i} $=$ 0 corresponds to a
pole-on view).  The best fit produces a dipole field strength, $B_d$
$=$ 9.5 MG, and a dipole offset, $a_z$ $=$ 0.06 (in units of stellar
radius).  The positive value of $a_z$ implies that the offset is {\it
toward} the observer.  Only $B_d$ is moderately constrained, both {\it
i} and $a_z$ can vary significantly yet still produce a reasonable fit
to the data \citep{1992ApJ...400..315B}.

{\bf WD 0310$-$624} is a DA WD that is one of the hottest in the new
sample.  Because of it's elevation significantly above the equal
temperature line (solid) in Figure \ref{teffcomp}, it is possible that
it is an unresolved double degenerate with very different component
effective temperatures.  In fact, this method has been used to
identify unresolved double degenerate candidates
\citep[i.e.][]{2001ApJS..133..413B}.


{\bf WD 0511$-$415} is a DA WD (spectrum is plotted in Figure
\ref{dahot}) whose spectral fit produces a $T_{\rm eff}$ $=$ 10,813
$\pm$ 219 K and a log $g$ $=$ 8.21 $\pm$ 0.10 using the spectral
fitting procedure of \citet{liebert03}.  This object lies near the red
edge of the ZZ Ceti instability strip as defined by
\citet{2006AJ....132..831G}.  If variable, this object would help to
constrain the cool edge of the instability strip in $T_{\rm eff}$, log
$g$ parameter space.  Follow-up high speed photometry is necessary to
confirm variability.

{\bf WD 0622$-$329} is a DAB WD displaying the Balmer lines as well as
weaker He {\scriptsize I} at 4472 and 5876 \AA.  The spectrum, shown
in Figure \ref{p5035}, is reproduced best with a model having $T_{\rm
eff}$ $\sim$43,700 K. However, the predicted He {\scriptsize II}
absorption line at 4686 \AA~for a WD of this $T_{\rm eff}$ is not
present in the spectrum.  In contrast, the SED fit to the photometry
implies a $T_{\rm eff}$ of $\sim$10,500 K (using either pure H or pure
He models).  Because the $T_{\rm eff}$ values are vastly discrepant,
we explored the possibility that this spectrum is not characterized by
a single temperature.  We modeled the spectrum assuming the object was
an unresolved double degenerate.  The best fit implies one component
is a DB with $T_{\rm eff}$ $=$ 14,170 $\pm$ 1,228 K and the other
component is a DA with $T_{\rm eff}$ $=$ 9,640 $\pm$ 303 K, similar
to the unresolved DA $+$ DB degenerate binary PG 1115+166 analyzed by
\citet{2002ApJ...566.1091B}.  One can see from Figure \ref{p5035} that
the spectrum is well modeled under this assumption.  We conclude this
object is likely a distant (well beyond 25 pc) unresolved double
degenerate.

{\bf WD 0840$-$136} is a DZ WD whose spectrum shows both Ca
{\scriptsize II} (H \& K) and Ca {\scriptsize I} (4226 \AA) lines as
shown in Figure \ref{p3327}.  Fits to the photometric data for
different atmospheric compositions indicate temperatures of about
4800-5000 K. However, fits to the optical spectrum using the models of
\citet{dufour07} cannot reproduce simultaneously all three calium
lines.  This problem is similar to that encountered by
\citet{dufour07} where the atmospheric parameters for the coolest DZ
WDs were considered uncertain because of possible high atmospheric
pressure effects.  We utilize a photometric relation relevant for WDs
of any atmospheric composition, which links $M_V$ to ($V - I$)
\citep{2004ApJ...601.1075S} to obtain a distance estimate of 19.3
$\pm$ 3.9 pc.

{\bf WD 1054$-$226} was observed spectroscopically as part of the
Edinburgh-Cape (EC) blue object survey and assigned a spectral type of
sdB+ \citep{1997MNRAS.287..867K}.  As is evident in Figure
\ref{dacool}, the spectrum of this object is the noisiest of all the
spectra presented here and perhaps a bit ambiguous.  As an additional
check, this object was recently observed using the ESO 3.6 m telescope
and has been confirmed to be a cool DA WD (Bergeron, private
communication).

{\bf WD 1105$-$340} is a DA WD (spectrum is plotted in Figure
\ref{dahot}) with a common proper motion companion with separation of
30$\farcs$6 at position angle 107.1$^\circ$.  The companion's spectral
type is M4Ve with $V_{J}$ $=$ 15.04, $R_{\rm KC}$ $=$ 13.68, $I_{\rm
KC}$ $=$ 11.96, $J$ $=$ 10.26, $H$ $=$ 9.70, and $K_S$ $=$ 9.41.  In
addition to the SED derived distance estimate for the WD, we utilize
the main sequence distance relations of \citet{2004AJ....128.2460H} to
estimate a distance to the red dwarf companion.  We obtain a distance
estimate of 19.1 $\pm$ 3.0 pc for the companion leaving open the
possibility that this system may lie just within 25 pc.  A
trigonometric parallax determination is currently underway for
confirmation.

{\bf WD 1149$-$272} is the only DQ WD discovered in the new sample.
This object was observed spectroscopically as part of the
Edinburgh-Cape (EC) blue object survey for which no features deeper
than 5\% were detected and was labeled a possible DC
\citep{1997MNRAS.287..867K}.  It is identified as having weak C$_2$
swan band absorption at 4737 and 5165 \AA~and is otherwise
featureless.  The DQ model reproduces the spectrum reliably and is
overplotted in Figure \ref{p4051}.  This object is characterized as
having $T_{\rm eff}$ $=$ 6188 $\pm$ 194 K and a log (C/He) $=$ $-$7.20
$\pm$ 0.16.

{\bf WD 2008$-$600} is a DC WD (spectrum is plotted in Figure
\ref{dc}) that is flux deficient in the near infrared, as indicated by
the 2MASS magnitudes.  The SED fit to the photometry is a poor match
to either the pure hydrogen or the pure helium models.  A pure
hydrogen model provides a slightly better match than a pure helium
model, and yields a $T_{\rm eff}$ of $\sim$3100 K, thereby placing it
in the relatively small sample of ultracool WDs.  In order to discern
the true nature of this object, we have constrained the model using
the distance obtained from the CTIOPI trigonometric parallax of 17.1
$\pm$ 0.4 pc (Subasavage et al., in preparation).  This object is then
best modeled as having mostly helium with trace amounts of hydrogen
(log (He/H) $=$ 2.61) in its atmosphere and has a $T_{\rm eff}$ $=$
5078 $\pm$ 221 K (see Figure \ref{scr2012}). A mixed hydrogen and
helium composition is required to produce sufficient absorption in the
infrared as a result of the collision-induced absorption by molecular
hydrogen due to collisions with helium.  Such mixed atmospheric
compositions have also been invoked to explain the infrared flux
deficiency in LHS 1126 \citep{1994ApJ...423..456B} as well as SDSS
1337+00 and LHS 3250 \citep{2002ApJ...580.1070B}.  While WD 2008$-$600
is likely {\it not} an ultracool WD, it is one of the brightest and
nearest cool WDs known. Because the 2MASS magnitudes are not very
reliable, we intend to obtain additional near-infrared photometry to
better constrain the fit.

{\bf WD 2138$-$332} is a DZ WD for which a calcium rich model
reproduces the spectrum reliably.  The spectrum and the overplotted
fit are shown in the bottom panel of Figure \ref{p3327}.  Clearly
evident in the spectrum are the strong Ca {\scriptsize II} absorption
at 3933 and 3968 \AA.  A weaker Ca {\scriptsize I} line is seen at
4226\AA.  Also seen are Mg {\scriptsize I} absorption lines at 3829,
3832, and 3838 \AA ~(blended) as well as Mg {\scriptsize I} at 5167,
5173, and 5184 \AA ~(also blended).  Several weak Fe {\scriptsize I}
lines from 4000\AA ~to 4500\AA ~and again from 5200\AA ~to 5500\AA
~are also present.  The divergence of the spectrum from the fit toward
the red end is likely due to an imperfect flux calibration of the
spectrum.  This object is characterized as having $T_{\rm eff}$ $=$
7188 $\pm$ 291 K and a log (Ca/He) $=$ $-$8.64 $\pm$ 0.16.  The
metallicity ratios are, at first, assumed to be solar \citep[as
defined by][]{1998SSRv...85..161G} and, in this case, the quality of
the fit was sufficient without deviation.  The corresponding log
(Mg/He) $=$ $-$7.42 $\pm$ 0.16 and log (Fe/He) $=$ $-$7.50 $\pm$ 0.16
for this object.

{\bf WD 2157$-$574} is a DA WD (spectrum is plotted in Figure
\ref{dacool}) unique to the new sample in that it displays weak Ca
{\scriptsize II} absorption at 3933 and 3968 \AA~(H and K) thereby
making its formal classification a DAZ.  Possible scenarios that
enrich the atmospheres of DAZs include accretion via (1) debris disks,
(2) ISM, and (3) cometary impacts (see \citealt*{2006ApJ...646..474K}
and references therein).  The 2MASS $K_S$ magnitude is near the faint
limit and is unreliable, but even considering the $J$ and $H$
magnitudes, there appears to be no appreciable near-infrared excess.
While this may tentatively rule out the possibility of a debris disk,
this object would be an excellent candidate for far-infrared
spaced-based studies to ascertain the origin of the enrichment.

\section {Discussion}

WDs represent the end state for stars less massive than $\sim$8
M$_\odot$ and are therefore relatively numerous.  Because of their
intrinsic faintness, only the nearby WD population can be easily
characterized and provides the benchmark upon which WD stellar
astrophysics is based.  It is clear from this work and others
(e.g.~\citealt*{2002ApJ...571..512H,2006ApJ...643..402K}) that the WD
sample is complete, at best, to only 13 pc.  Spectroscopic
confirmation of new WDs as well as trigonometric parallax
determinations for both new and known WDs will lead to a more complete
sample and will push the boundary of completeness outward.  We
estimate that 8 new WDs and an additional 12 known WDs without
trigonometric parallaxes are nearer than 25 pc, including one within
10 pc (WD 0141$-$675).  Parallax measurements via CTIOPI are underway
for these 20 objects to confirm proximity.  This total of 20 WDs
within 25 pc constitutes an 18\% increase to the 109 WDs with
trigonometric parallaxes $\ge$ 40 mas.

Evaluating the proper motions of the new and known samples within 25
pc indicates that almost double the number of systems have been found
with $\mu$ $<$ 1$\farcs$0 yr$^{-1}$ than with $\mu$ $\ge$ 1$\farcs$0
yr$^{-1}$ (13 vs 7, see Table \ref{diststats}).  The only WD estimated
to be within 10 pc has $\mu$ $>$ 1$\farcs$0 yr$^{-1}$, although WD
1202$-$232 is estimated to be 10.2 $\pm$ 1.7 pc and it's proper motion
is small ($\mu$ $=$ 0$\farcs$227 yr$^{-1}$).

Because this effort focuses mainly on the southern hemisphere, it is
likely that there is a significant fraction of nearby WDs in the
northern hemisphere that have also gone undetected.  With the recent
release of the LSPM-North Catalog \citep{2005AJ....129.1483L}, these
objects are identifiable by employing the same techniques used in this
work.  The challenge is the need for a large scale parallax survey
focusing on WDs to confirm proximity.  Since the {\it HIPPARCOS}
mission, only six WD trigonometric parallaxes have been published
\citep{1999MNRAS.309L..33H,2003A&A...404..317S}, and of those, only
two are within 25 pc.  The USNO parallax program is in the process of
publishing trigonometric parallaxes for $\sim$130 WDs, mostly in the
northern hemisphere, although proximity was not a primary motivation
for target selection (Dahn, private communication).

In addition to further completing the nearby WD census, the wealth of
observational data available from this effort provides reliable
constraints on their physical parameters (i.e.~$T_{\rm eff}$, log $g$,
mass, and radius).  Unusual objects are then revealed, such as
those discussed in $\S$ 4.2.  In particular, trigonometric parallaxes
help identify WDs that are overluminous, as is the case for WD
0121$-$429.  This object, and others similar to it, are excellent
candidates to provide insight into binary evolution.  If they can be
resolved using high resolution astrometric techniques (i.e.~speckle,
adaptive optics, or interferometry via {\it Hubble Space Telescope's}
Fine Guidance Sensors), they may provide astrometric masses, which are
fundamental calibrators for stellar structure theory and for the
reliability of the theoretical WD mass-radius and
initial-to-final-mass relationships.  To date, only four WD
astrometric masses are known to better than $\sim$ 5\%
\citep{1998ApJ...494..759P}.


One avenue that is completely unexplored to date is a careful high
resolution search for planets around WDs.  Theory dictates that the
Sun will become a WD, and when it does, the outer planets will remain
in orbit (not without transformations of their own, of course).  In
this scenario, the Sun will have lost more than half of its mass,
thereby amplifying the signature induced by the planets.  Presumably,
this has already occurred in the Milky Way and systems such as these
merely await detection.  Because of the faintness and spectral
signatures of WDs (i.e.~few, if any, broad absorption lines), current
radial velocity techniques are inadequate for planet detection,
leaving astrometric techniques as the only viable option. For a given
system, the astrometric signature is inversely related to distance
(i.e.~the nearer the system, the larger the astrometric signature).
This effort aims to provide a complete census of nearby WDs that can
be probed for these astrometric signatures using future astrometric
efforts.

\section {Acknowledgments}

The RECONS team at Georgia State University wishes to thank the NSF
(grant AST 05-07711), NASA's Space Interferometry Mission, and GSU for
their continued support of our study of nearby stars.  We also thank
the continuing support of the members of the SMARTS consortium, who
enable the operations of the small telescopes at CTIO where all of the
data in this work were collected.  J.~P.~S.~is indebted to Wei-Chun
Jao for the use of his photometry reduction pipeline.  P.~B.~is a
Cottrell Scholar of Research Corporation and would like to thank the
NSERC Canada for its support.  N.~C.~H.~would like to thank colleagues
in the Wide Field Astronomy Unit at Edinburgh for their efforts
contributing to the existence of the SSS; particular thanks go to Mike
Read, Sue Tritton, and Harvey MacGillivray.  This work has made use of
the SIMBAD, VizieR, and Aladin databases, operated at the CDS in
Strasbourg, France.  We have also used data products from the Two
Micron All Sky Survey, which is a joint project of the University of
Massachusetts and the Infrared Processing and Analysis Center, funded
by NASA and NSF.

\appendix

\section{Appendix}

In order to ensure correct cross-referencing of names for the new and
known WD systems presented here, Table \ref{alternate} lists
additional names found in the literature.  Objects for which there is
an NLTT designation will also have the corresponding L or LP
designations found in the NLTT catalog.  This is necessary because the
NLTT designations were not published in the original catalog, but
rather are the record numbers in the electronic version of the catalog
and have been adopted out of necessity.


\clearpage
%

\voffset+0pt{
\begin{deluxetable}{lrrrcrrrrrrr@{$\pm$}rlr@{$\pm$}lll}
\rotate \tabletypesize{\tiny} \tablecaption{Optical and Infrared
Photometry, and Derived Parameters for New and Known White Dwarfs.
\label{photometry}} 
\tablewidth{0pt} \tablehead{\colhead{WD}        &
			    \colhead{$V_J$}     &
			    \colhead{$R_C$}     &
			    \colhead{$I_C$}     &
			    \colhead{\#}        &
			    \colhead{$J$}       &
			    \colhead{$\sigma_J$}&
			    \colhead{$H$}       &
			    \colhead{$\sigma_H$}&
			    \colhead{$K_S$}     &
			    \colhead{$\sigma_{K_S}$}&
                            \multicolumn{2}{c}{$T_{\rm eff}$}   &   
                            \colhead{Comp}             &
			    \multicolumn{2}{c}{Dist}       &
			    \colhead{SpT}       &
			    \colhead{Notes}     \\

			    \colhead{Name}       &
			    \colhead{}           &
			    \colhead{}           &
			    \colhead{}           &
			    \colhead{Obs}        &
			    \colhead{}           &
			    \colhead{}           &
			    \colhead{}           &
			    \colhead{}           &
			    \colhead{}           &
			    \colhead{}           &
                            \multicolumn{2}{c}{(K)}    &
			    \colhead{}           &
			    \multicolumn{2}{c}{(pc)}   &
			    \colhead{}           &
			    \colhead{}}

\startdata

\tableline
\vspace{-6pt}\\
\multicolumn{18}{c}{New Spectroscopically Confirmed White Dwarfs} \\
\vspace{-6pt}\\
\tableline
\vspace{-5pt}\\

0034$-$602.............  &  14.08   &  14.19   &  14.20   &   3   &  14.37  &  0.04  &  14.55  &  0.06  &  14.52  &  0.09  &  14655  &   1413  &  H       &   35.8  &   5.7   & DA3.5  &                   \\
0121$-$429.............  &  14.83   &  14.52   &  14.19   &   4   &  13.85  &  0.02  &  13.63  &  0.04  &  13.53  &  0.04  &   6369  &   137   &  H       & \nodata & \nodata & DAH    &  \tablenotemark{a}\\
0216$-$398.............  &  15.75   &  15.55   &  15.29   &   3   &  15.09  &  0.04  &  14.83  &  0.06  &  14.89  &  0.14  &   7364  &   241   &  H       &   29.9  &   4.7   & DA7.0  &                   \\
0253$-$755.............  &  16.70   &  16.39   &  16.08   &   2   &  15.77  &  0.07  &  15.76  &  0.15  &  15.34  &  null  &   6235  &   253   &  He      &   34.7  &   5.5   & DC     &                   \\
0310$-$624.............  &  15.92   &  15.99   &  16.03   &   2   &  16.13  &  0.10  &  16.31  &  0.27  &  16.50  &  null  &  13906  &   1876  &  H       & \nodata & \nodata & DA3.5  &  \tablenotemark{b}\\
0344$+$014.............  &  16.52   &  16.00   &  15.54   &   2   &  15.00  &  0.04  &  14.87  &  0.09  &  14.70  &  0.12  &   5084  &    91   &  He      &   19.9  &   3.1   & DC     &                   \\
0404$-$510.............  &  15.81   &  15.76   &  15.70   &   2   &  15.74  &  0.06  &  15.55  &  0.13  &  15.59  &  null  &  10052  &   461   &  H       &   53.5  &   8.5   & DA5.0  &                   \\
0501$-$555.............  &  16.35   &  16.17   &  15.98   &   2   &  15.91  &  0.08  &  15.72  &  0.15  &  15.82  &  0.26  &   7851  &   452   &  He      &   44.8  &   6.9   & DC     &                   \\
0511$-$415.............  &  16.00   &  15.99   &  15.93   &   2   &  15.96  &  0.08  &  15.97  &  0.15  &  15.20  &  null  &  10393  &   560   &  H       &   61.8  &  10.8   & DA5.0  &                   \\
0525$-$311.............  &  15.94   &  16.03   &  16.03   &   2   &  16.20  &  0.12  &  16.21  &  0.25  &  14.98  &  null  &  12941  &   1505  &  H       &   76.3  &  13.6   & DA4.0  &                   \\
0607$-$530.............  &  15.99   &  15.92   &  15.78   &   3   &  15.82  &  0.07  &  15.66  &  0.14  &  15.56  &  0.21  &   9395  &   426   &  H       &   51.7  &   9.0   & DA5.5  &                   \\
0622$-$329.............  &  15.47   &  15.41   &  15.36   &   2   &  15.44  &  0.06  &  15.35  &  0.11  &  15.53  &  0.25  & \nodata & \nodata & \nodata  & \nodata & \nodata & DAB    &  \tablenotemark{c}\\
0821$-$669.............  &  15.34   &  14.82   &  14.32   &   3   &  13.79  &  0.03  &  13.57  &  0.03  &  13.34  &  0.04  &   5160  &    95   &  H       &   11.5  &   1.9   & DA10.0 &                   \\
0840$-$136.............  &  15.72   &  15.36   &  15.02   &   3   &  14.62  &  0.03  &  14.42  &  0.05  &  14.54  &  0.09  & \nodata & \nodata & \nodata  & \nodata & \nodata & DZ     &  \tablenotemark{d}\\
1016$-$308.............  &  14.67   &  14.75   &  14.81   &   2   &  15.05  &  0.04  &  15.12  &  0.08  &  15.41  &  0.21  &  16167  &   1598  &  H       &   50.6  &   9.2   & DA3.0  &                   \\
1054$-$226.............  &  16.02   &  15.82   &  15.62   &   2   &  15.52  &  0.05  &  15.40  &  0.11  &  15.94  &  0.26  &   8266  &   324   &  H       &   41.0  &   7.0   & DA6.0  &  \tablenotemark{e}\\
1105$-$340.............  &  13.66   &  13.72   &  13.79   &   2   &  13.95  &  0.03  &  13.98  &  0.04  &  14.05  &  0.07  &  13926  &   988   &  H       &   28.2  &   4.8   & DA3.5  &  \tablenotemark{f}\\  
1149$-$272.............  &  15.87   &  15.59   &  15.37   &   4   &  15.17  &  0.05  &  14.92  &  0.06  &  14.77  &  0.11  &   6188  &   194   &  He (+C) &   24.0  &   3.8   & DQ     &                   \\
1243$-$123.............  &  15.57   &  15.61   &  15.64   &   2   &  15.74  &  0.07  &  15.73  &  0.11  &  16.13  &  null  &  12608  &   1267  &  H       &   62.6  &  10.7   & DA4.0  &                   \\
1316$-$215.............  &  16.67   &  16.33   &  15.99   &   2   &  15.56  &  0.05  &  15.33  &  0.08  &  15.09  &  0.14  &   6083  &   201   &  H       &   31.6  &   5.3   & DA8.5  &                   \\
1436$-$781.............  &  16.11   &  15.82   &  15.49   &   2   &  15.04  &  0.04  &  14.88  &  0.08  &  14.76  &  0.14  &   6246  &   200   &  H       &   26.0  &   4.3   & DA8.0  &                   \\
1452$-$310.............  &  15.85   &  15.77   &  15.63   &   2   &  15.58  &  0.06  &  15.54  &  0.09  &  15.50  &  0.22  &   9206  &   375   &  H       &   46.8  &   8.1   & DA5.5  &                   \\
1647$-$327.............  &  16.21   &  15.85   &  15.49   &   3   &  15.15  &  0.05  &  14.82  &  0.08  &  14.76  &  0.11  &   6092  &   193   &  H       &   25.5  &   4.2   & DA8.5  &                   \\
1742$-$722.............  &  15.53   &  15.62   &  15.70   &   2   &  15.85  &  0.08  &  15.99  &  0.18  &  15.65  &  null  &  15102  &   2451  &  H       &   71.7  &  12.9   & DA3.5  &                   \\
1946$-$273.............  &  14.19   &  14.31   &  14.47   &   2   &  14.72  &  0.04  &  14.77  &  0.09  &  14.90  &  0.13  &  21788  &   3304  &  H       &   52.0  &   9.9   & DA2.5  &                   \\
2008$-$600.............  &  15.84   &  15.40   &  14.99   &   4   &  14.93  &  0.05  &  15.23  &  0.11  &  15.41  &  null  &   5078  &   221   &  He      & \nodata & \nodata & DC     &  \tablenotemark{g}\\
2008$-$799.............  &  16.35   &  15.96   &  15.57   &   3   &  15.11  &  0.04  &  15.03  &  0.08  &  14.64  &  0.09  &   5807  &   161   &  H       &   24.5  &   4.1   & DA8.5  &                   \\
2035$-$369.............  &  14.94   &  14.85   &  14.72   &   2   &  14.75  &  0.04  &  14.72  &  0.06  &  14.84  &  0.09  &   9640  &   298   &  H       &   33.1  &   5.7   & DA5.0  &                   \\
2103$-$397.............  &  15.31   &  15.15   &  14.91   &   2   &  14.79  &  0.03  &  14.63  &  0.04  &  14.64  &  0.08  &   7986  &   210   &  H       &   28.2  &   4.8   & DA6.5  &                   \\
2138$-$332.............  &  14.47   &  14.30   &  14.16   &   3   &  14.17  &  0.03  &  14.08  &  0.04  &  13.95  &  0.06  &   7188  &   291   &  He (+Ca)&   17.3  &   2.7   & DZ     &                   \\
2157$-$574.............  &  15.96   &  15.73   &  15.49   &   3   &  15.18  &  0.04  &  15.05  &  0.07  &  15.28  &  0.17  &   7220  &   246   &  H       &   32.0  &   5.4   & DAZ    &                   \\
2218$-$416.............  &  15.36   &  15.35   &  15.24   &   2   &  15.38  &  0.04  &  15.14  &  0.09  &  15.39  &  0.15  &  10357  &   414   &  H       &   45.6  &   8.0   & DA5.0  &                   \\
2231$-$387.............  &  16.02   &  15.88   &  15.62   &   2   &  15.57  &  0.06  &  15.51  &  0.11  &  15.11  &  0.15  &   8155  &   336   &  H       &   40.6  &   6.9   & DA6.0  &                   \\

\vspace{-5pt}\\
\tableline
\vspace{-6pt}\\
\multicolumn{18}{c}{Known White Dwarfs without a Trigonometric Parallax Estimated to be Within 25 pc} \\
\vspace{-6pt}\\
\tableline
\vspace{-5pt}\\

0141$-$675 ............   &  13.82   &  13.52   &  13.23   &   3   &  12.87  &  0.02  &  12.66  &  0.03  &  12.58  &  0.03  &  6484 &  128 & H   &  9.7  &  1.6  &  DA8.0   &                   \\
0806$-$661 ............   &  13.73   &  13.66   &  13.61   &   3   &  13.70  &  0.02  &  13.74  &  0.03  &  13.78  &  0.04  & 10753 &  406 & He  & 21.1  &  3.5  &  DQ      &                   \\
1009$-$184 ............   &  15.44   &  15.18   &  14.91   &   3   &  14.68  &  0.04  &  14.52  &  0.05  &  14.31  &  0.07  &  6449 &  194 & He  & 20.9  &  3.2  &  DZ      &  \tablenotemark{h}\\
1036$-$204 ............   &  16.24   &  15.54   &  15.34   &   3   &  14.63  &  0.03  &  14.35  &  0.04  &  14.03  &  0.07  &  4948 &   70 & He  & 16.2  &  2.5  &  DQ      &  \tablenotemark{i}\\
1202$-$232 ............   &  12.80   &  12.66   &  12.52   &   3   &  12.40  &  0.02  &  12.30  &  0.03  &  12.34  &  0.03  &  8623 &  168 & H   & 10.2  &  1.7  &  DA6.0   &                   \\
1315$-$781 ............   &  16.16   &  15.73   &  15.35   &   2   &  14.89  &  0.04  &  14.67  &  0.08  &  14.58  &  0.12  &  5720 &  162 & H   & 21.6  &  3.6  &  DC      &  \tablenotemark{j}\\
1339$-$340 ............   &  16.43   &  16.00   &  15.56   &   2   &  15.00  &  0.04  &  14.75  &  0.06  &  14.65  &  0.10  &  5361 &  138 & H   & 21.2  &  3.5  &  DA9.5   &                   \\
1756$+$143 ............   &  16.30   &  16.12   &  15.69   &   1   &  14.93  &  0.04  &  14.66  &  0.06  &  14.66  &  0.08  &  5466 &  151 & H   & 22.4  &  3.4  &  DA9.0   &  \tablenotemark{k}\\
1814$+$134 ............   &  15.85   &  15.34   &  14.86   &   2   &  14.38  &  0.04  &  14.10  &  0.06  &  14.07  &  0.06  &  5313 &  115 & H   & 15.6  &  2.5  &  DA9.5   &                   \\
2040$-$392 ............   &  13.74   &  13.77   &  13.68   &   2   &  13.77  &  0.02  &  13.82  &  0.03  &  13.81  &  0.05  & 10811 &  325 & H   & 23.1  &  4.0  &  DA4.5   &                   \\
2211$-$392 ............   &  15.91   &  15.61   &  15.24   &   2   &  14.89  &  0.03  &  14.64  &  0.05  &  14.56  &  0.08  &  6243 &  167 & H   & 23.5  &  4.0  &  DA8.0   &                   \\
2226$-$754A...........    &  16.57   &  15.93   &  15.33   &   2   &  14.66  &  0.04  &  14.66  &  0.06  &  14.44  &  0.08  &  4230 &  104 & H   & 12.8  &  2.0  &  DC      &  \tablenotemark{l}\\
2226$-$754B...........    &  16.88   &  16.17   &  15.51   &   2   &  14.86  &  0.04  &  14.82  &  0.06  &  14.72  &  0.12  &  4177 &  112 & H   & 14.0  &  2.2  &  DC      &  \tablenotemark{l}\\

\vspace{-5pt}\\
\tableline
\vspace{-6pt}\\
\multicolumn{18}{c}{Known White Dwarfs without a Trigonometric Parallax Estimated to be Beyond 25 pc} \\
\vspace{-6pt}\\
\tableline
\vspace{-5pt}\\
 
0024$-$556.............   &  15.17   &  15.15   &  15.07   &   2   &  15.01  &  0.04  &  15.23  &  0.10  &  15.09  &  0.14  &  10007  &   378   &  H      &   39.8  &   6.8   &   DA5.0  &                   \\
0150$+$256.............   &  15.70   &  15.52   &  15.33   &   2   &  15.07  &  0.04  &  15.07  &  0.09  &  15.15  &  0.14  &   7880  &   280   &  H      &   33.0  &   5.6   &   DA6.5  &                   \\
0255$-$705 ............   &  14.08   &  14.03   &  14.00   &   2   &  14.04  &  0.03  &  14.12  &  0.04  &  13.99  &  0.06  &  10541  &   326   &  H      &   25.8  &   4.5   &   DA5.0  &                   \\
0442$-$304.............   &  16.03   &  15.93   &  15.86   &   2   &  15.94  &  0.09  &  15.81  &  null  &  15.21  &  null  &   9949  &   782   &  He     &   55.1  &   9.1   &   DQ     &                   \\
0928$-$713 ............   &  15.11   &  14.97   &  14.83   &   3   &  14.77  &  0.03  &  14.69  &  0.06  &  14.68  &  0.09  &   8836  &   255   &  H      &   30.7  &   5.3   &   DA5.5  &                   \\
1143$-$013.............   &  16.39   &  16.08   &  15.79   &   1   &  15.54  &  0.06  &  15.38  &  0.08  &  15.18  &  0.16  &   6824  &   250   &  H      &   34.4  &   5.8   &   DA7.5  &                   \\
1237$-$230 ............   &  16.53   &  16.13   &  15.74   &   2   &  15.35  &  0.05  &  15.08  &  0.08  &  14.94  &  0.11  &   5841  &   173   &  H      &   26.9  &   4.5   &   DA8.5  &                   \\
1314$-$153.............   &  14.82   &  14.89   &  14.97   &   2   &  15.17  &  0.05  &  15.26  &  0.09  &  15.32  &  0.21  &  15604  &  2225   &  H      &   52.7  &   9.5   &   DA3.0  &                   \\
1418$-$088 ............   &  15.39   &  15.21   &  15.01   &   2   &  14.76  &  0.04  &  14.73  &  0.06  &  14.76  &  0.10  &   7872  &   243   &  H      &   28.5  &   4.8   &   DA6.5  &                   \\
1447$-$190.............   &  15.80   &  15.59   &  15.32   &   2   &  15.06  &  0.04  &  14.87  &  0.07  &  14.78  &  0.11  &   7153  &   235   &  H      &   29.1  &   4.9   &   DA7.0  &                   \\
1607$-$250.............   &  15.19   &  15.12   &  15.09   &   2   &  15.08  &  0.08  &  15.08  &  0.08  &  15.22  &  0.15  &  10241  &   457   &  H      &   41.2  &   7.2   &   DA5.0  &                   \\

\enddata

\tablenotetext{a}{Distance via SED fit (not listed) is underestimated because object is likely an unresolved double degenerate with one magnetic component (see $\S$ 4.2).  Instead, we adopt the trigonometric parallax distance of 17.7 $\pm$ 0.7 pc derived via CTIOPI.}
\tablenotetext{b}{Distance via SED fit (not listed) is underestimated because object is likely a distant (well beyond 25 pc) unresolved double degenerate (see $\S$ 4.2).}
\tablenotetext{c}{Distance via SED fit (not listed) is underestimated because object is likely a distant (well beyond 25 pc) unresolved double degenerate with components of type DA and DB (see $\S$ 4.2).  Temperatures derived from the spectroscopic fit yield 9,640 $\pm$ 303 K and 14,170 $\pm$ 1,228 K for the DA and DB respectively.}
\tablenotetext{d}{Object is likely cooler than $T_{\rm eff}$ $\sim$5000 K and the theoretical models do not provide an accurate treatment at these temperatures (see $\S$ 4.2).  Instead, we use the linear photometric distance relation of \citet{2004ApJ...601.1075S} and obtain a distance estimate of 19.3 $\pm$ 3.9 pc.}
\tablenotetext{e}{This object was observed as part of the Edinburgh-Cape survey and was classified as a sdB+ \citep{1997MNRAS.287..867K}.}
\tablenotetext{f}{Distance of 19.1 $\pm$ 3.0 pc is estimated using $VRIJHK_S$ for the common proper motion companion M dwarf and the relations of \citet{2004AJ....128.2460H}.  System is possibly within 25 pc. (see $\S$ 4.2).} 
\tablenotetext{g}{Distance estimate is undetermined.  Instead, we adopt the distance measured via trogonometric parallax of 17.1 $\pm$ 0.4 pc (see $\S$ 4.2).}
\tablenotetext{h}{Not listed in \citet{1999ApJS..121....1M} but identified as a DC/DQ WD by \citet{2002AJ....123.2002H}.  We obtained blue spectra that show Ca II H \& K absorption and classify this object as a DZ.}
\tablenotetext{i}{The SED fit to the photometry is marginal.  This object displays deep swan band absorption that significantly affects its measured magnitudes.}
\tablenotetext{j}{Not listed in \citet{1999ApJS..121....1M} but identified as a WD by \citet{1949ApJ...109..528L}.  Spectral type is derived from our spectra.}
\tablenotetext{k}{As of mid-2004, object has moved onto a background source.  Photometry is probably contaminated, which is consistent with the poor SED fit for this object.}
\tablenotetext{l}{Spectral type was determined using spectra published by \citet{2002ApJ...565..539S}.}

\end{deluxetable}

%
\begin{deluxetable}{lccc}
\tabletypesize{\footnotesize}
\tablecaption{Distance Estimate Statistics for New and Known White Dwarfs. 
\label{diststats}}
\tablewidth{0pt}

\tablehead{\vspace{-15pt} \\
           \colhead{Proper motion}&
           \colhead{d $\leq$ 10 pc}&
           \colhead{10 pc $<$ d $\leq$ 25 pc}&
           \colhead{d $>$ 25 pc}}

\startdata
$\mu$ $\geq$ 1$\farcs$0 yr$^{-1}$.........................      &  1  &  6  &   1  \\
1$\farcs$0 yr$^{-1}$ $>$ $\mu$ $\geq$ 0$\farcs$8 yr$^{-1}$......&  0  &  0  &   0  \\
0$\farcs$8 yr$^{-1}$ $>$ $\mu$ $\geq$ 0$\farcs$6 yr$^{-1}$......&  0  &  2  &   2  \\
0$\farcs$6 yr$^{-1}$ $>$ $\mu$ $\geq$ 0$\farcs$4 yr$^{-1}$......&  0  &  6  &  11  \\
0$\farcs$4 yr$^{-1}$ $>$ $\mu$ $\geq$ 0$\farcs$18 yr$^{-1}$.... &  0  &  5  &  22  \\
{    }Total....................................                 &  1  & 19  &  36  \\
\enddata

\end{deluxetable}

\begin{deluxetable}{lccccll@{ $=$ }l@{ $=$ }l}
\tabletypesize{\tiny} \tablecaption{Astrometry and Alternate Designations for New and Known White Dwarfs.
\label{alternate}} 
\tablewidth{0pt} 
\tablehead{\colhead{WD Name}                      &
           \colhead{RA}        &
           \colhead{Dec}       &
           \colhead{PM}        &
           \colhead{PA}        &
           \colhead{Ref}       &
           \multicolumn{3}{c}{Alternate Names}     \\

           \colhead{}           &
           \colhead{(J2000.0)}  &
           \colhead{(J2000.0)}  &
           \colhead{(arcsec yr$^{-1}$)}&
           \colhead{(deg)} &
           \colhead{}           &
	   \colhead{}           &
	   \colhead{}           &
	   \colhead{}}

\startdata   
	     
\tableline
\vspace{-6pt}\\
\multicolumn{9}{c}{New Spectroscopically Confirmed White Dwarfs} \\
\vspace{-6pt}\\
\tableline
\vspace{-5pt}\\

0034$-$602.........  &  00 36 22.31  &  $-$59 55 27.5  &  0.280  &  069.0  &  L  &  NLTT 1993      &  LP 122-4     &  \nodata       \\
0121$-$429.........  &  01 24 03.98  &  $-$42 40 38.5  &  0.538  &  155.2  &  L  &  LHS 1243       &  NLTT 4684    &  LP 991-16     \\
0216$-$398.........  &  02 18 31.51  &  $-$39 36 33.2  &  0.500  &  078.6  &  L  &  LHS 1385       &  NLTT 7640    &  LP 992-99     \\
0253$-$755.........  &  02 52 45.64  &  $-$75 22 44.5  &  0.496  &  063.5  &  S  &  SCR 0252-7522  &  \nodata      &  \nodata       \\
0310$-$624.........  &  03 11 21.34  &  $-$62 15 15.7  &  0.416  &  083.3  &  S  &  SCR 0311-6215  &  \nodata      &  \nodata       \\
0344$+$014.........  &  03 47 06.82  &  $+$01 38 47.5  &  0.473  &  150.4  &  S  &  LHS 5084       &  NLTT 11839   &  LP 593-56     \\ 
0404$-$510.........  &  04 05 32.86  &  $-$50 55 57.8  &  0.320  &  090.7  &  P  &  LEHPM 1-3634   &  \nodata      &  \nodata       \\
0501$-$555.........  &  05 02 43.43  &  $-$55 26 35.2  &  0.280  &  191.9  &  P  &  LEHPM 1-3865   &  \nodata      &  \nodata       \\
0511$-$415.........  &  05 13 27.80  &  $-$41 27 51.7  &  0.292  &  004.4  &  P  &  LEHPM 2-1180   &  \nodata      &  \nodata       \\
0525$-$311.........  &  05 27 24.33  &  $-$31 06 55.7  &  0.379  &  200.7  &  P  &  NLTT 15117     &  LP 892-45    &  LEHPM 2-521   \\ 
0607$-$530.........  &  06 08 43.81  &  $-$53 01 34.1  &  0.246  &  327.6  &  P  &  LEHPM 2-2008   &  \nodata      &  \nodata       \\
0622$-$329.........  &  06 24 25.78  &  $-$32 57 27.4  &  0.187  &  177.7  &  P  &  LEHPM 2-5035   &  \nodata      &  \nodata       \\
0821$-$669.........  &  08 21 26.70  &  $-$67 03 20.1  &  0.758  &  327.6  &  S  &  SCR 0821-6703  &  \nodata      &  \nodata       \\
0840$-$136.........  &  08 42 48.45  &  $-$13 47 13.1  &  0.272  &  263.0  &  S  &  NLTT 20107     &  LP 726-1     &  \nodata       \\
1016$-$308.........  &  10 18 39.84  &  $-$31 08 02.0  &  0.212  &  304.0  &  L  &  NLTT 23992     &  LP 904-3     &  LEHPM 2-5779  \\
1054$-$226.........  &  10 56 38.64  &  $-$22 52 55.9  &  0.277  &  349.7  &  P  &  NLTT 25792     &  LP 849-31    &  LEHPM 2-1372  \\
1105$-$340.........  &  11 07 47.89  &  $-$34 20 51.4  &  0.287  &  168.0  &  S  &  SCR 1107-3420A &  \nodata      &  \nodata       \\ 
1149$-$272.........  &  11 51 36.10  &  $-$27 32 21.0  &  0.199  &  278.3  &  P  &  LEHPM 2-4051   &  \nodata      &  \nodata       \\
1243$-$123.........  &  12 46 00.69  &  $-$12 36 19.9  &  0.406  &  305.4  &  S  &  SCR 1246-1236  &  \nodata      &  \nodata       \\
1316$-$215.........  &  13 19 24.72  &  $-$21 47 55.0  &  0.467  &  179.2  &  S  &  NLTT 33669     &  LP 854-50    &  WT 2034       \\
1436$-$781.........  &  14 42 51.54  &  $-$78 23 53.6  &  0.409  &  272.0  &  S  &  NLTT 38003     &  LP 40-109    &  LTT 5814      \\
1452$-$310.........  &  14 55 23.47  &  $-$31 17 06.4  &  0.199  &  174.2  &  P  &  LEHPM 2-4029   &  \nodata      &  \nodata       \\
1647$-$327.........  &  16 50 44.32  &  $-$32 49 23.2  &  0.526  &  193.8  &  L  &  LHS 3245       &  NLTT 43628   &  LP 919-1      \\
1742$-$722.........  &  17 48 31.21  &  $-$72 17 18.5  &  0.294  &  228.2  &  P  &  LEHPM 2-1166   &  \nodata      &  \nodata       \\
1946$-$273.........  &  19 49 19.78  &  $-$27 12 25.7  &  0.213  &  162.0  &  L  &  NLTT 48270     &  LP 925-53    &  \nodata       \\
2008$-$600.........  &  20 12 31.75  &  $-$59 56 51.5  &  1.440  &  165.6  &  S  &  SCR 2012-5956  &  \nodata      &  \nodata       \\
2008$-$799.........  &  20 16 49.66  &  $-$79 45 53.0  &  0.434  &  128.4  &  S  &  SCR 2016-7945  &  \nodata      &  \nodata       \\
2035$-$369.........  &  20 38 41.42  &  $-$36 49 13.5  &  0.230  &  104.0  &  L  &  NLTT 49589     &  L 495-42     &  LEHPM 2-3290  \\
2103$-$397.........  &  21 06 32.01  &  $-$39 35 56.7  &  0.266  &  151.7  &  P  &  LEHPM 2-1571   &  \nodata      &  \nodata       \\
2138$-$332.........  &  21 41 57.56  &  $-$33 00 29.8  &  0.210  &  228.5  &  P  &  NLTT 51844     &  L 570-26     &  LEHPM 2-3327  \\
2157$-$574.........  &  22 00 45.37  &  $-$57 11 23.4  &  0.233  &  252.0  &  P  &  LEHPM 1-4327   &  \nodata      &  \nodata       \\
2218$-$416.........  &  22 21 25.37  &  $-$41 25 27.0  &  0.210  &  143.4  &  P  &  LEHPM 1-4598   &  \nodata      &  \nodata       \\
2231$-$387.........  &  22 33 54.47  &  $-$38 32 36.9  &  0.370  &  220.5  &  P  &  NLTT 54169     &  LP 1033-28   &  LEHPM 1-4859  \\

\vspace{-5pt}\\
\tableline
\vspace{-6pt}\\
\multicolumn{9}{c}{Known White Dwarfs without a Trigonometric Parallax Estimated to be Within 25 pc} \\
\vspace{-6pt}\\
\tableline
\vspace{-5pt}\\

0141$-$675 ........  &  01 43 00.98  &  $-$67 18 30.3  &  1.048  &  197.8  &  L  &  LHS 145         &  NLTT 5777    &  L 88-59      \\ 
0806$-$661 ........  &  08 06 53.76  &  $-$66 18 16.6  &  0.454  &  131.4  &  S  &  NLTT 19008      &  L 97-3       &  \nodata      \\
1009$-$184 ........  &  10 12 01.88  &  $-$18 43 33.2  &  0.519  &  268.2  &  S  &  WT 1759         &  LEHPM 2-220  &  \nodata      \\
1036$-$204 ........  &  10 38 55.57  &  $-$20 40 56.7  &  0.628  &  330.3  &  L  &  LHS 2293        &  NLTT 24944   &  LP 790-29    \\
1202$-$232 ........  &  12 05 26.66  &  $-$23 33 12.1  &  0.227  &  002.0  &  L  &  NLTT 29555      &  LP 852-7     &  LEHPM 2-1894 \\
1315$-$781 ........  &  13 19 25.63  &  $-$78 23 28.3  &  0.477  &  139.2  &  S  &  NLTT 33551      &  L 40-116     &  \nodata      \\
1339$-$340 ........  &  13 42 02.88  &  $-$34 15 19.4  &  2.547  &  296.7  &  Le &  PM J13420-3415  &  \nodata      &  \nodata      \\
1756$+$143 ........  &  17 58 22.90  &  $+$14 17 37.8  &  1.014  &  235.4  &  Le &  LSR 1758+1417   &  \nodata      &  \nodata      \\  
1814$+$134 ........  &  18 17 06.48  &  $+$13 28 25.0  &  1.207  &  201.5  &  Le &  LSR 1817+1328   &  \nodata      &  \nodata      \\
2040$-$392 ........  &  20 43 49.21  &  $-$39 03 18.0  &  0.306  &  179.0  &  L  &  NLTT 49752      &  L 495-82     &  \nodata      \\
2211$-$392 ........  &  22 14 34.75  &  $-$38 59 07.3  &  1.056  &  110.1  &  O  &  WD J2214-390    &  LEHPM 1-4466 &  \nodata      \\
2226$-$754A........  &  22 30 40.00  &  $-$75 13 55.3  &  1.868  &  167.5  &  S  &  SSSPM J2231-7514&  \nodata      &  \nodata      \\  
2226$-$754B........  &  22 30 33.55  &  $-$75 15 24.2  &  1.868  &  167.5  &  S  &  SSSPM J2231-7515&  \nodata      &  \nodata      \\

\vspace{-5pt}\\
\tableline
\vspace{-6pt}\\
\multicolumn{9}{c}{Known White Dwarfs without a Trigonometric Parallax Estimated to be Beyond 25 pc} \\
\vspace{-6pt}\\
\tableline
\vspace{-5pt}\\

0024$-$556.........  &  00 26 40.69  &  $-$55 24 44.1  &  0.580  &  211.8  &  L  &  LHS 1076       &  NLTT 1415    &  L 170-27      \\
0150$+$256.........  &  01 52 51.93  &  $+$25 53 40.7  &  0.220  &  076.0  &  L  &  NLTT 6275      &  G 94-21      &  \nodata       \\
0255$-$705.........  &  02 56 17.22  &  $-$70 22 10.8  &  0.682  &  097.9  &  L  &  LHS 1474       &  NLTT 9485    &  L 54-5        \\
0442$-$304.........  &  04 44 29.38  &  $-$30 21 14.2  &  0.196  &  199.5  &  P  &  NLTT 13882     &  LP 891-65    &  HE 0442-3027  \\
0928$-$713.........  &  09 29 07.97  &  $-$71 33 58.8  &  0.439  &  320.2  &  S  &  NLTT 21957     &  L 64-40      &  \nodata       \\
1143$-$013.........  &  11 46 25.77  &  $-$01 36 36.8  &  0.563  &  140.2  &  S  &  LHS 2455       &  NLTT 28493   &  \nodata       \\ 
1237$-$230.........  &  12 40 24.18  &  $-$23 17 43.8  &  1.102  &  219.9  &  L  &  LHS 339        &  NLTT 31473   &  LP 853-15     \\
1314$-$153.........  &  13 16 43.59  &  $-$15 35 58.3  &  0.708  &  196.7  &  L  &  LHS 2712       &  NLTT 33503   &  LP 737-47     \\
1418$-$088.........  &  14 20 54.93  &  $-$09 05 08.7  &  0.480  &  266.8  &  S  &  LHS 5270       &  NLTT 37026   &  \nodata       \\
1447$-$190.........  &  14 50 11.93  &  $-$19 14 08.7  &  0.253  &  285.4  &  P  &  NLTT 38499     &  LP 801-14    &  LEHPM 2-1835  \\
1607$-$250.........  &  16 10 50.21  &  $-$25 13 16.0  &  0.209  &  314.0  &  L  &  NLTT 42153     &  LP 861-31    &  \nodata       \\

\enddata

\tablenotetext{~}{References. ---
(L) ~\citealt{1979lccs.book.....L, nltt},
(Le)~\citealt{2003AJ....126..921L}, ~\citealt{2005ApJ...633L.121L},
(O) ~\citealt{2001Sci...292..698O},
(P) ~\citealt{2004A&A...421..763P},
(S) ~\citealt{2005AJ....129..413S, 2005AJ....130.1658S}, this work
}

\end{deluxetable}

\clearpage

%
\figcaption[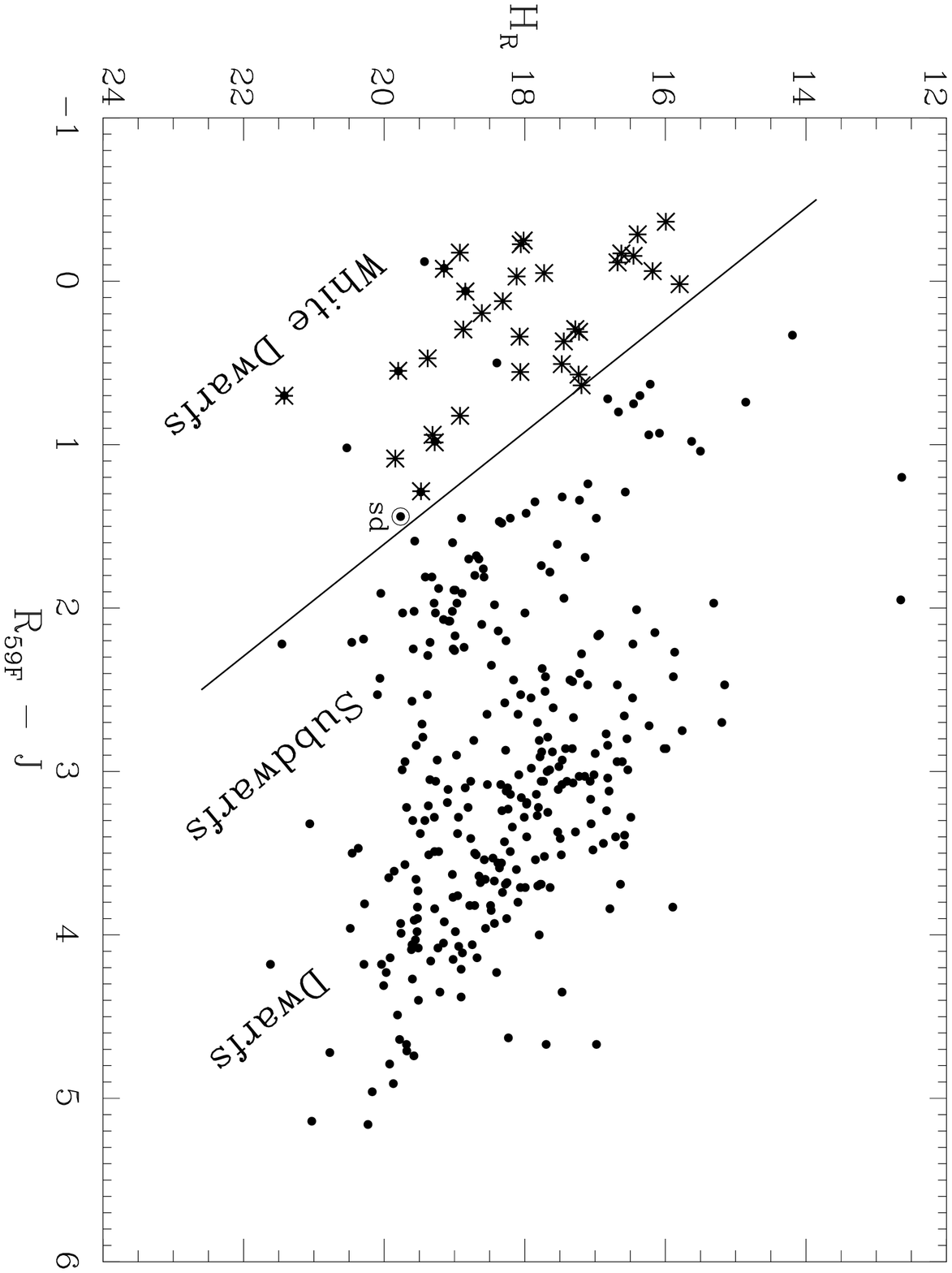]{Reduced proper motion diagram used to select
WD candidates for spectroscopic follow-up.  Plotted are the new high
proper motion objects from
\citet{2005AJ....129..413S,2005AJ....130.1658S}.  The line is a
somewhat arbitrary boundary between the WDs (below) and the subdwarfs
(just above).  Main sequence dwarfs fall above and to the right of the
subdwarfs, although there is significant overlap.  Asterisks indicate
the 33 new WDs reported here.  Three dots in the WD region are
deferred to a future paper.  The point labeled ``sd'' is a confirmed
subdwarf contaminant of the WD sample.
\label{rpm}}



\figcaption[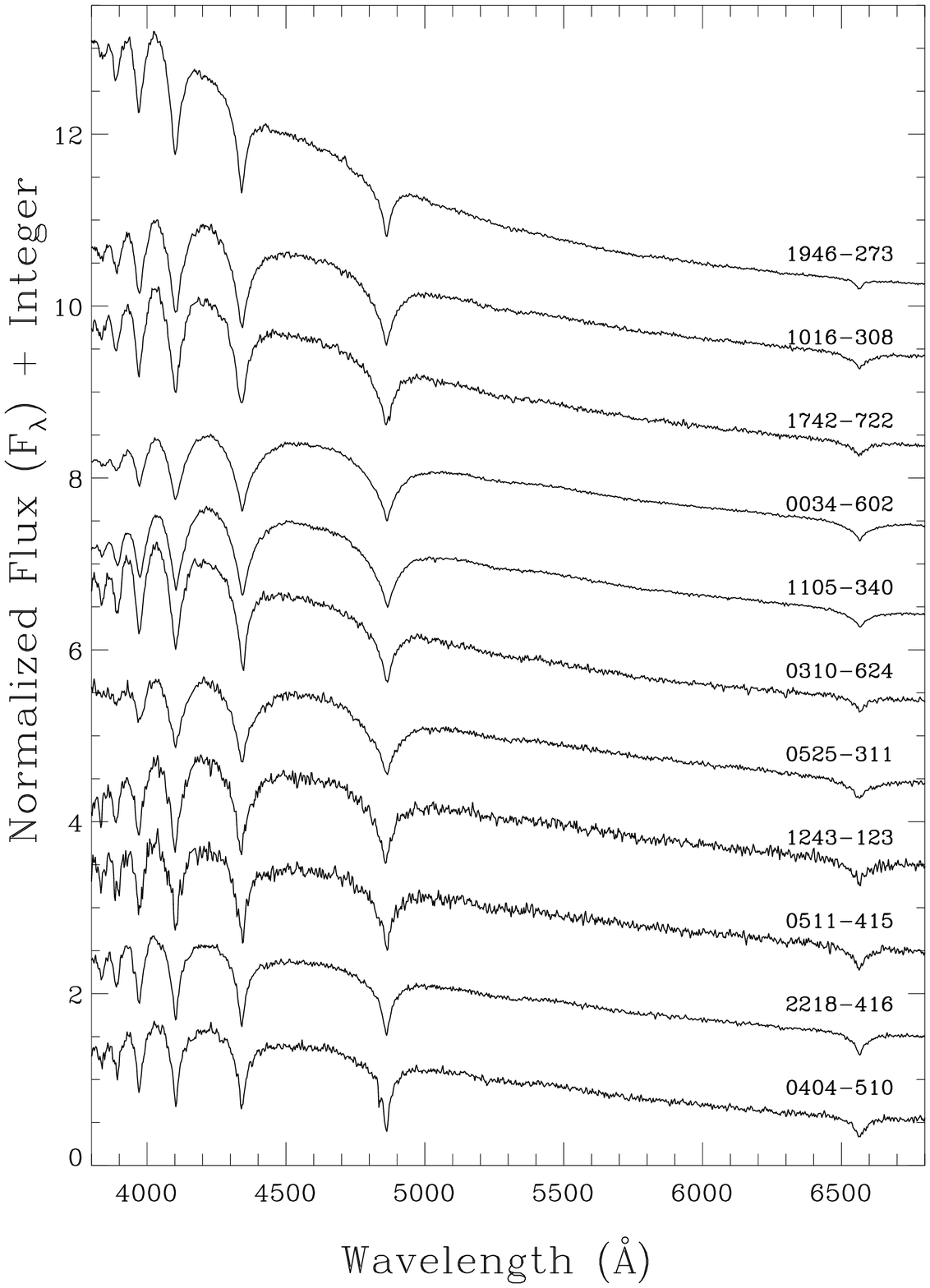]{Spectral plots of the hot ($T_{\rm eff}$ $\ge$
10000 K) DA WDs from the new sample, plotted in descending $T_{\rm
eff}$ as derived from the SED fits to the photometry.  Note that some
of the flux calibrations are not perfect, in particular, at the blue
end.
\label{dahot}}

\figcaption[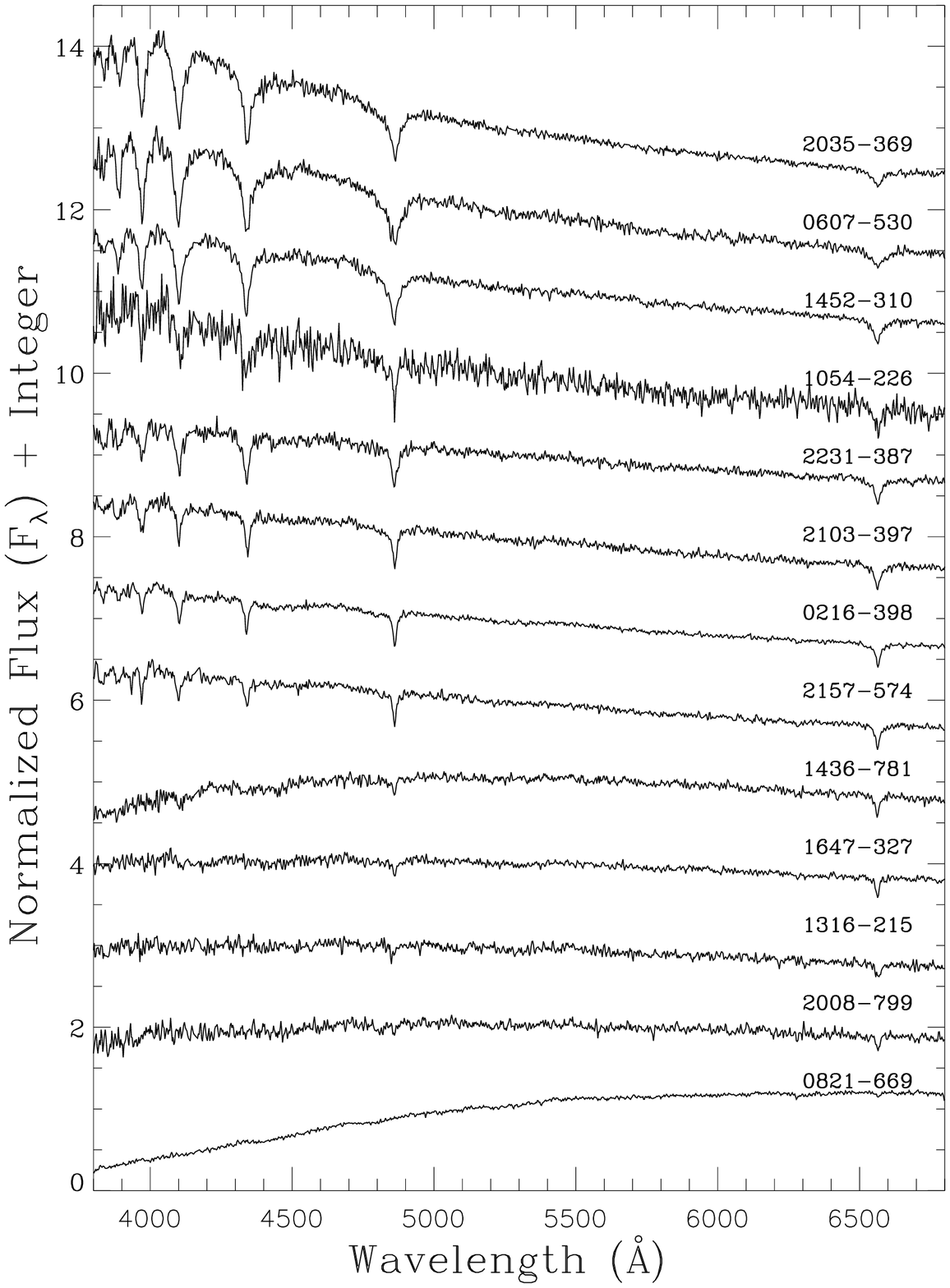]{Spectral plots of cool ($T_{\rm eff}$ $<$ 10000
K) DA WDs from the new sample, plotted in descending $T_{\rm eff}$ as
derived from the SED fits to the photometry.  Note that some of the
flux calibrations are not perfect, in particular, at the blue end.
\label{dacool}}

\figcaption[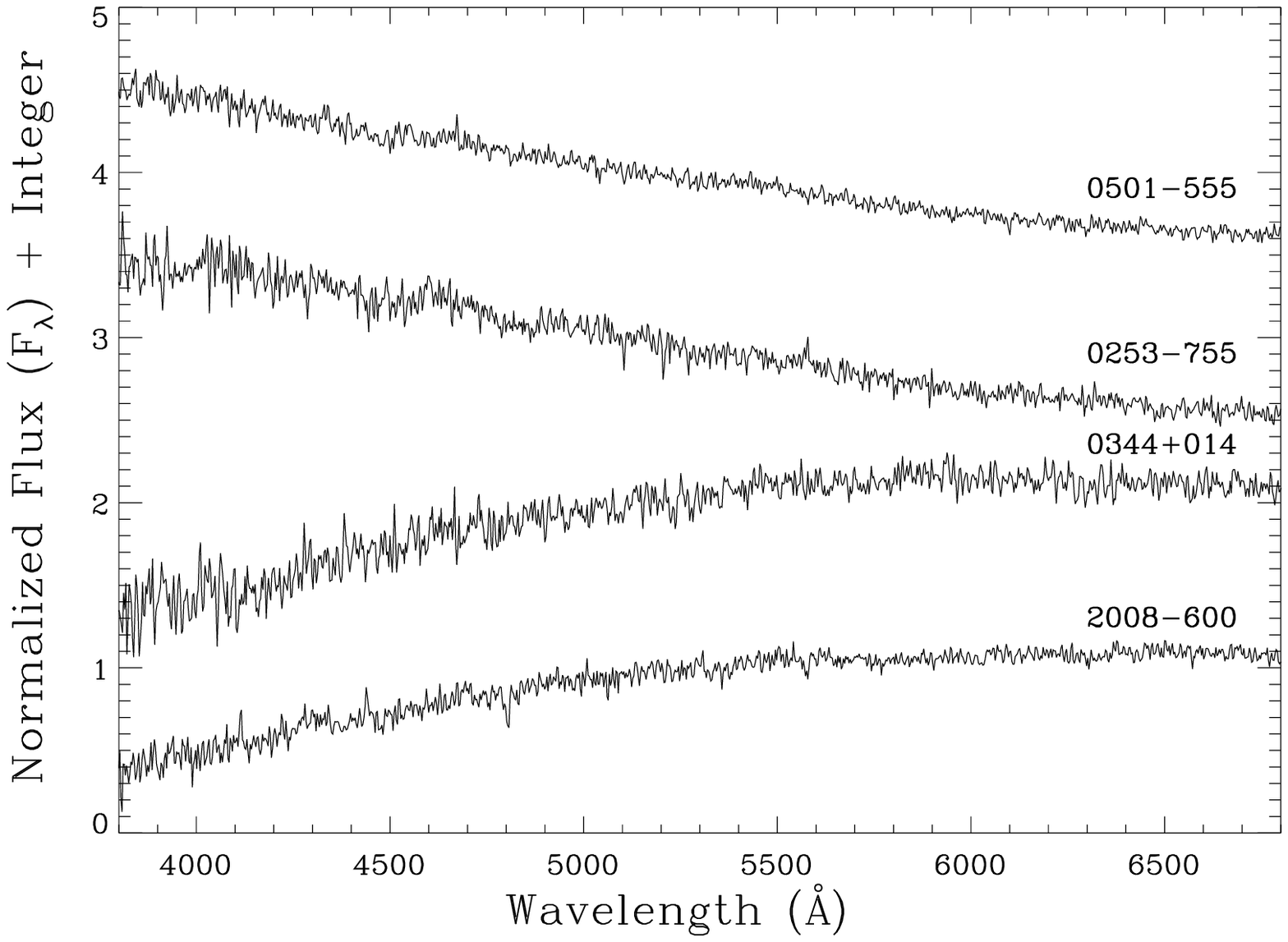]{Spectral plots of the four featureless DC white
dwarfs from the new sample, plotted in descending $T_{\rm eff}$ as
derived from the SED fits to the photometry.  Note that some of the
flux calibrations are not perfect, in particular, at the blue end.
\label{dc}}

\figcaption[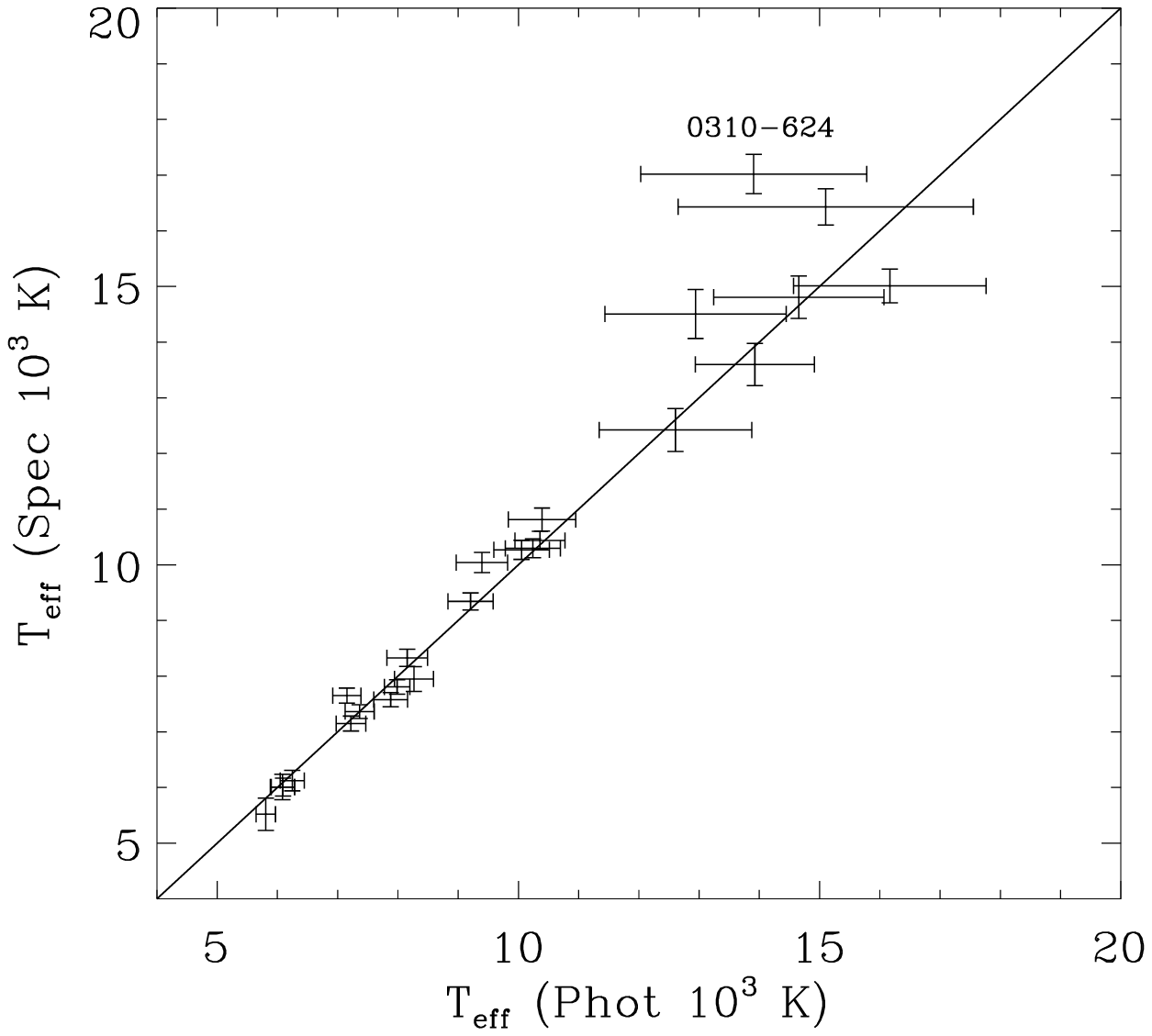]{Comparison plot of the values of $T_{\rm
eff}$ derived from photometric SED fitting vs those derived from
spectral fitting for 25 of the DA WDs in the new sample.  The solid
line represents equal temperatures.  The elevated point, 0310$-$624,
is discussed in $\S$ 4.2.
\label{teffcomp}}

\figcaption[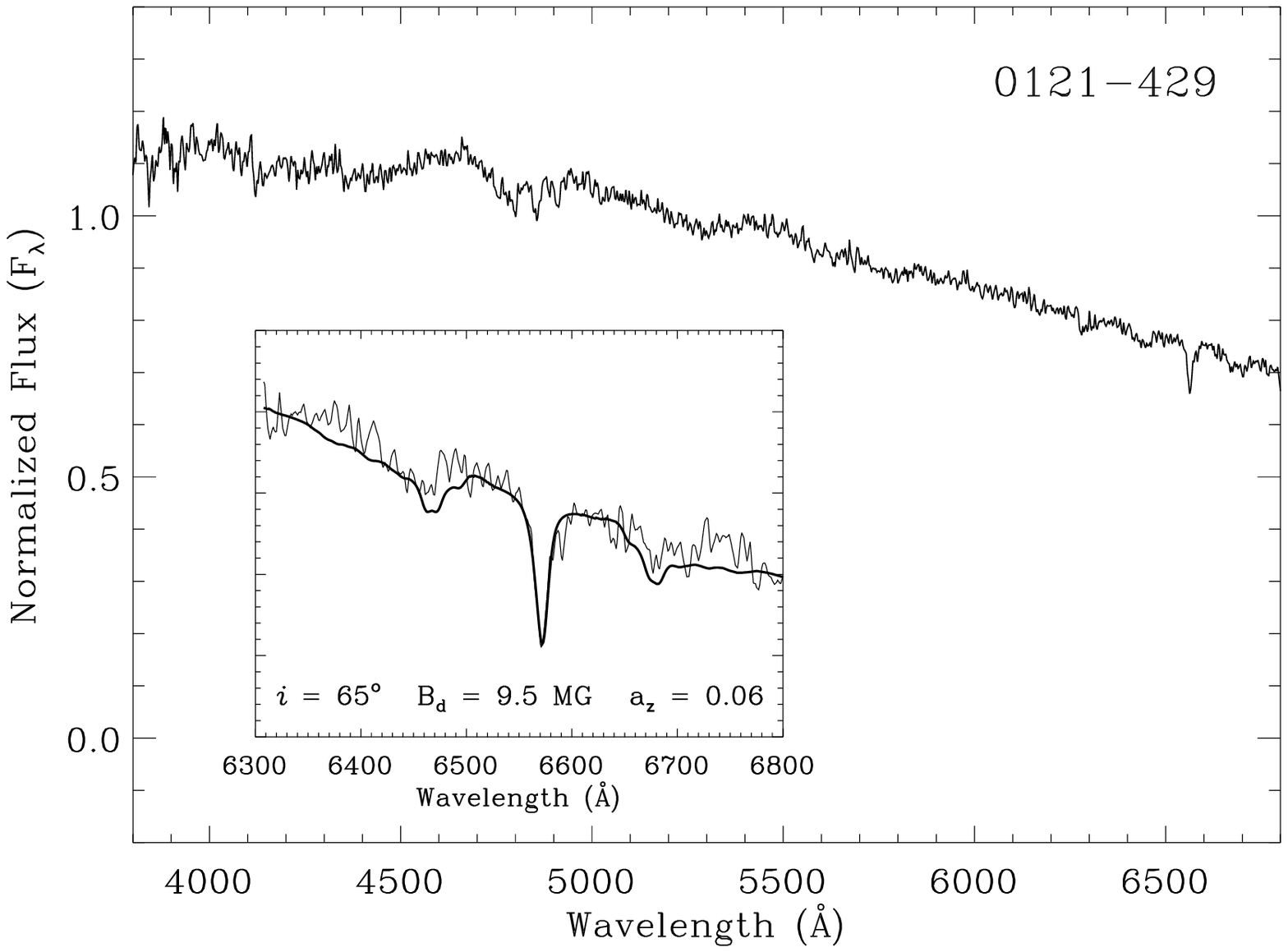]{Spectral plot of WD 0121$-$429.  The inset
plot displays the spectrum (light line) in the H$\alpha$ region to
which a magnetic fit (heavy line), as outlined in
\citet{1992ApJ...400..315B}, was performed using the $T_{\rm eff}$
obtained from the SED fit to the photometry.  The resulting magnetic
parameters are listed below the fit.
\label{lhs1243}}

\figcaption[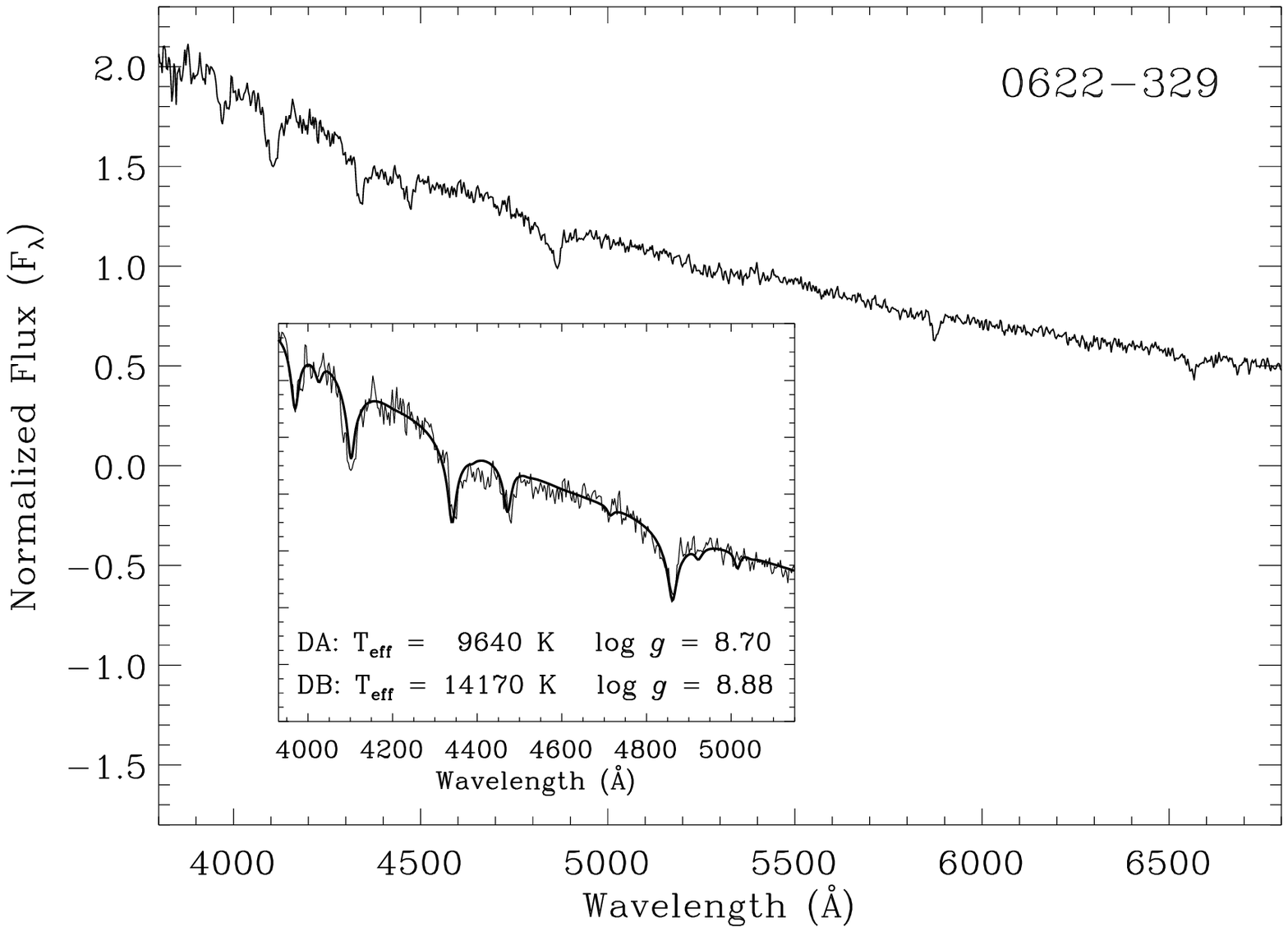]{Spectral plot of WD 0622$-$329.  The inset plot
displays the spectrum (light line) in the region to which the model
(heavy line) was fit assuming the spectrum is a convolution of a DB
component and a slightly cooler DA component.  Best fit physical
parameters are listed below the fit for each component.
\label{p5035}}

\figcaption[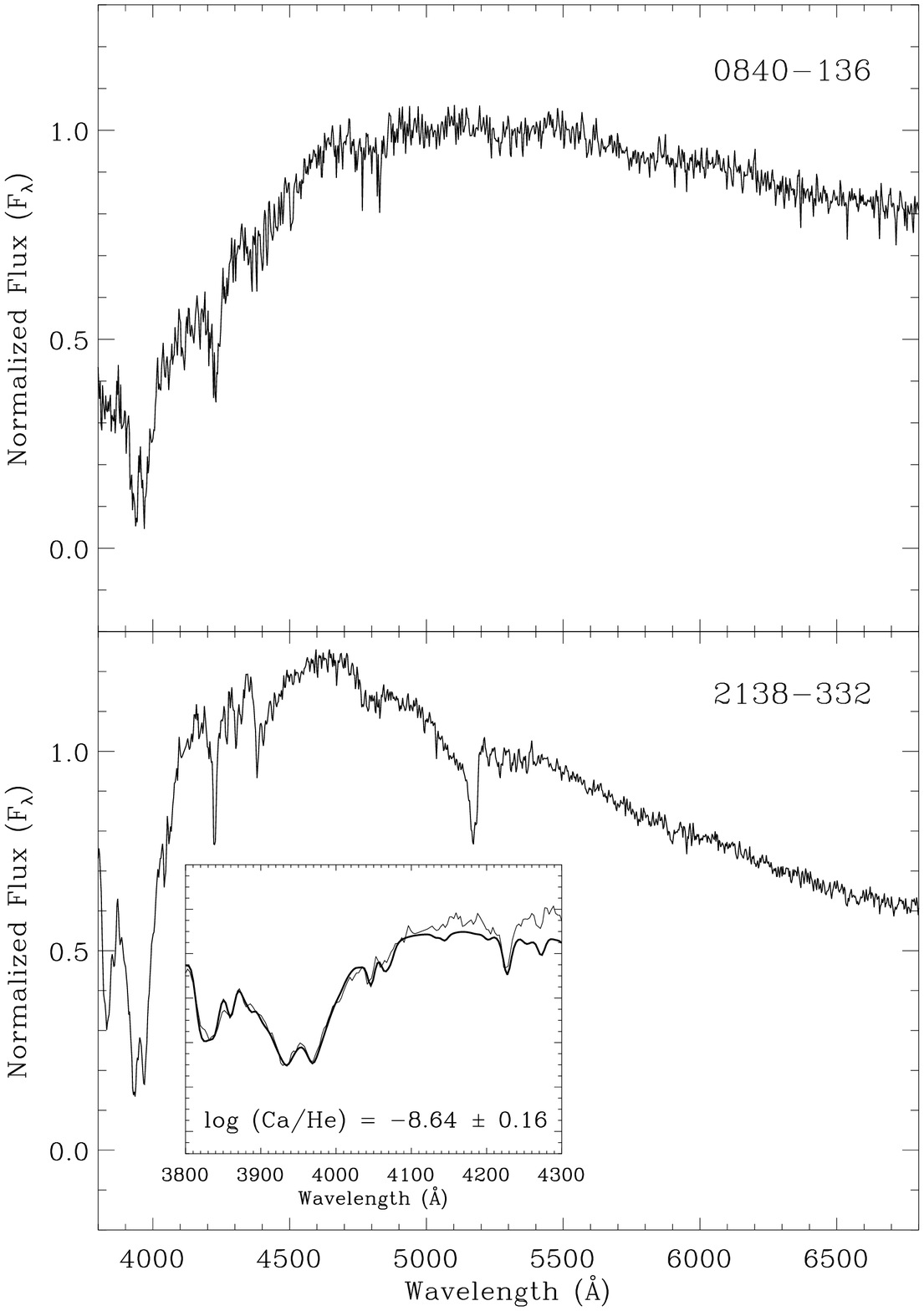]{({\it top panel}) Spectral plot of WD
0840$-$136.  The DZ model failed to reproduce the spectrum presumably
because this object is cooler than $T_{\rm eff}$ $\sim$ 5000 K where
additional pressure effects, not included in the model, become
important.  ({\it bottom panel}) Spectral plot of WD 2138$-$332.  The
inset plot displays the spectrum (light line) in the region to which
the model (heavy line) was fit.
\label{p3327}}

\figcaption[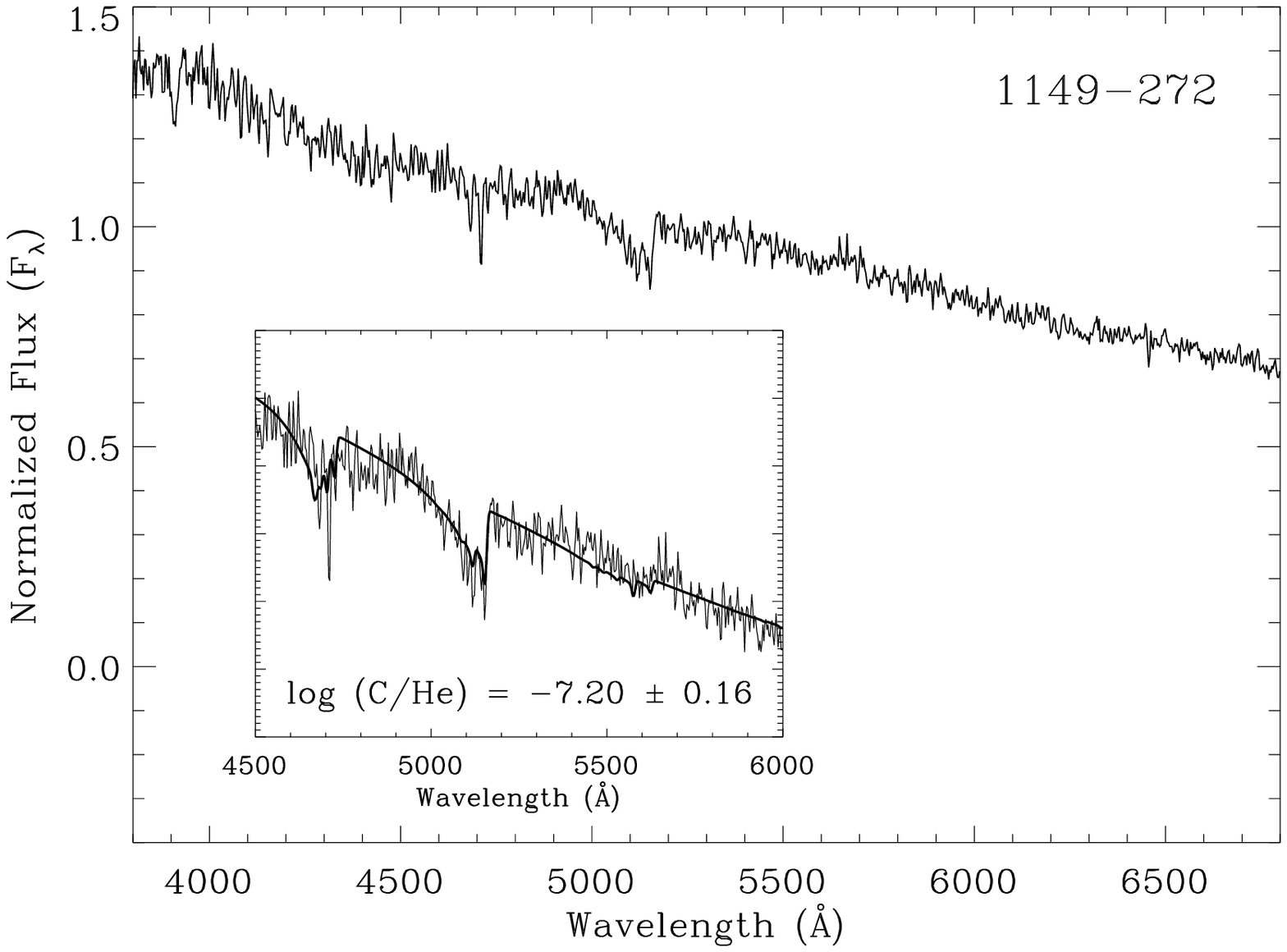]{Spectral plot of WD 1149$-$272.  The inset plot
displays the spectrum (light line) in the region to which the model
(heavy line) was fit.
\label{p4051}}


\figcaption[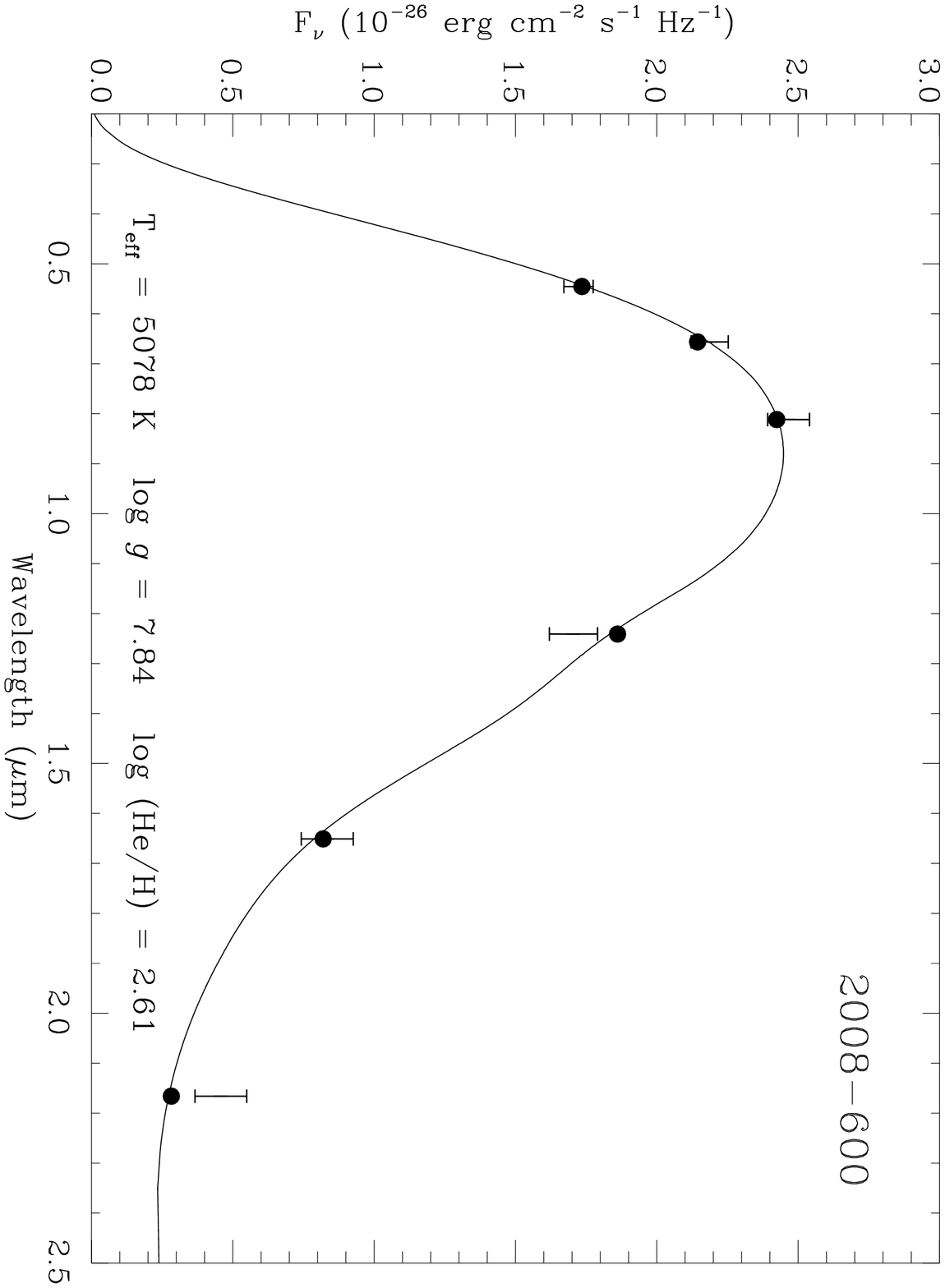]{Spectral energy distribution plot of WD
2008$-$600 with the distance constrained by the trigonometric distance
of 17.1 $\pm$ 0.4 pc.  Best fit physical parameters are listed below
the fit.  Points are fit values; error bars are derived from the
uncertainties in the magnitudes and the parallax.
\label{scr2012}}

\begin{figure}
\plotone{f1.eps}
\end{figure}

\clearpage

%

%

\begin{figure}
\plotone{f2.eps}
\end{figure}

\clearpage

\begin{figure}
\plotone{f3.eps}
\end{figure}

\clearpage

\begin{figure}
\plotone{f4.eps}
\end{figure}

\clearpage

\begin{figure}
\plotone{f5.eps}
\end{figure}

\clearpage

\begin{figure}
\plotone{f6.eps}
\end{figure}

\clearpage

\begin{figure}
\plotone{f7.eps}
\end{figure}

\clearpage

\begin{figure}
\plotone{f8.eps}
\end{figure}

\clearpage

\begin{figure}
\plotone{f9.eps}
\end{figure}

\clearpage

\begin{figure}
\plotone{f10.eps}
\end{figure}


\begin{thebibliography}{}

\bibitem[Bergeron et al.(1992a)]{1992ApJ...400..315B} Bergeron, P., Ruiz, 
M.-T., \& Leggett, S.~K.\ 1992, \apj, 400, 315 

\bibitem[Bergeron et al.(1992b)]{1992ApJ...394..228B} Bergeron, P., Saffer, 
R.~A., \& Liebert, J.\ 1992, \apj, 394, 228 

\bibitem[Bergeron et al.(1993)]{1993ApJ...407..733B} Bergeron, P., Ruiz, 
M.-T., \& Leggett, S.~K.\ 1993, \apj, 407, 733 

\bibitem[Bergeron et al.(1994)]{1994ApJ...423..456B} Bergeron, P., Ruiz, 
M.-T., Leggett, S.~K., Saumon, D., \& Wesemael, F.\ 1994, \apj, 423, 456 

\bibitem[Bergeron et al.(1997)]{BRL} Bergeron, P., Ruiz, M. T., \& Leggett, 
S. K. 1997, \apjs, 108, 339

\bibitem[Bergeron et al.(2001)]{2001ApJS..133..413B} Bergeron, P., Leggett, 
S.~K., \& Ruiz, M.~T.\ 2001, \apjs, 133, 413 
 
\bibitem[Bergeron \& Leggett(2002)]{2002ApJ...580.1070B} Bergeron, P., \& 
Leggett, S.~K.\ 2002, \apj, 580, 1070 

\bibitem[Bergeron \& Liebert(2002)]{2002ApJ...566.1091B} Bergeron, P., \& 
Liebert, J.\ 2002, \apj, 566, 1091 

\bibitem[Bessel(1990)]{1990A&AS...83..357B} Bessel, M.~S.\ 1990, \aaps, 83, 
357 

\bibitem[Carpenter(2001)]{2001AJ....121.2851C} Carpenter, J.~M.\ 2001, \aj, 
121, 2851 

\bibitem[Dufour et al.(2005)]{dufour05} Dufour, P., Bergeron, P., \& 
Fontaine, G.\ 2005, \apj, 627, 404

\bibitem[Dufour et al.(2007)]{dufour07} Dufour, P., Bergeron, P., 
Liebert, J., Harris, H.~C., Knapp, G.~R., Anderson, S.~F., Hall, P.~B.,
Strauss, M.~A., Collinge, M.~J., \& Edwards, M.~C. 2007, submitted to ApJ

\bibitem[Filippenko(1982)]{1982PASP...94..715F} Filippenko, A.~V.\ 1982, 
\pasp, 94, 715

\bibitem[Finch et al.(2007)]{finch} Finch, C.~T., Henry, T.~J.,
Subasavage, J.~P., Jao, W.-C., Hambly, N.~C.\ 2007, \aj, submitted

\bibitem[Fontaine et al.(2001)]{fon01} Fontaine, G., Brassard, P., \& 
Bergeron, P. 2001, \pasp, 113, 409

\bibitem[Gianninas et al.(2006)]{2006AJ....132..831G} Gianninas, A., 
Bergeron, P., \& Fontaine, G.\ 2006, \aj, 132, 831 

\bibitem[Gliese \& Jahrei{\ss}(1991)]{1991adc..rept.....G} Gliese, W., \& 
Jahrei{\ss}, H.\ 1991, On: The Astronomical Data Center CD-ROM: Selected 
Astronomical Catalogs, Vol.~I; L.E.~Brotzmann, S.E.~Gesser (eds.), 
NASA/Astronomical Data Center, Goddard Space Flight Center, Greenbelt, MD 

\bibitem[Graham(1982)]{1982PASP...94..244G} Graham, J.~A.\ 1982, \pasp, 94, 
244 

\bibitem[Grevesse \& Sauval(1998)]{1998SSRv...85..161G} Grevesse, N., \& 
Sauval, A.~J.\ 1998, Space Science Reviews, 85, 161 

\bibitem[Hambly et al.(1999)]{1999MNRAS.309L..33H} Hambly, N.~C., Smartt, 
S.~J., Hodgkin, S.~T., Jameson, R.~F., Kemp, S.~N., Rolleston, W.~R.~J., \& 
Steele, I.~A.\ 1999, \mnras, 309, L33 

\bibitem[Henry et al.(2002)]{2002AJ....123.2002H} Henry, T.~J., Walkowicz, 
L.~M., Barto, T.~C., \& Golimowski, D.~A.\ 2002, \aj, 123, 2002 

\bibitem[Henry et al.(2003)]{2003fst3.book..111H} Henry, T.~J., Backman, 
D.~E., Blackwell, J., Okimura, T., \& Jue, S.\ 2003, The Future of Small 
Telescopes In The New Millennium.~Volume III - Science in the Shadow of 
Giants, 111

\bibitem[Henry et al.(2004)]{2004AJ....128.2460H} Henry, T.~J., Subasavage, 
J.~P., Brown, M.~A., Beaulieu, T.~D., Jao, W., \& Hambly, N.~C.\ 2004, \aj, 
128, 2460 

\bibitem[Henry et al.(2006)]{2006AJ....132.2360H} Henry, T.~J., Jao, W.-C., 
Subasavage, J.~P., Beaulieu, T.~D., Ianna, P.~A., Costa, E., \& M{\'e}ndez, 
R.~A.\ 2006, \aj, 132, 2360 

\bibitem[Holberg et al.(2002)]{2002ApJ...571..512H} Holberg, J.~B., Oswalt, 
T.~D., \& Sion, E.~M.\ 2002, \apj, 571, 512 

\bibitem[Holberg et al.(2006)]{holberg06} Holberg, J. B., \& Bergeron, P. 
2006, \aj, 132, 1223

\bibitem[Iben \& Renzini(1984)]{1984PhR...105..329I} Iben, I., \& Renzini, 
A.\ 1984, \physrep, 105, 329

\bibitem[Jao et al.(2005)]{2005AJ....129.1954J} Jao, W.-C., Henry, T.~J., 
Subasavage, J.~P., Brown, M.~A., Ianna, P.~A., Bartlett, J.~L., Costa, E., 
\& M{\'e}ndez, R.~A.\ 2005, \aj, 129, 1954 

\bibitem[Kawka \& Vennes(2006)]{2006ApJ...643..402K} Kawka, A., \& Vennes, 
S.\ 2006, \apj, 643, 402 

\bibitem[Kilic et al.(2006)]{2006ApJ...646..474K} Kilic, M., von Hippel, 
T., Leggett, S.~K., \& Winget, D.~E.\ 2006, \apj, 646, 474 

\bibitem[Kilkenny et al.(1997)]{1997MNRAS.287..867K} Kilkenny, D., 
O'Donoghue, D., Koen, C., Stobie, R.~S., \& Chen, A.\ 1997, \mnras, 287, 
867 

\bibitem[Kleinman et al.(2004)]{2004ApJ...607..426K} Kleinman, S.~J., et 
al.\ 2004, \apj, 607, 426 

\bibitem[Landolt(1992)]{1992AJ....104..340L} Landolt, A.~U.\ 1992,
\aj, 104, 340
 
\bibitem[L{\' e}pine et al.(2003)]{2003AJ....126..921L} L{\'
e}pine, S., Shara, M.~M., \& Rich, R.~M.\ 2003, \aj, 126, 921

\bibitem[L{\'e}pine \& Shara(2005)]{2005AJ....129.1483L} L{\'e}pine, S., \& 
Shara, M.~M.\ 2005, \aj, 129, 1483 

\bibitem[L{\'e}pine et al.(2005)]{2005ApJ...633L.121L} L{\'e}pine, S., 
Rich, R.~M., \& Shara, M.~M.\ 2005, \apjl, 633, L121 

\bibitem[Liebert et al.(2003)]{liebert03} Liebert, J., Bergeron, P.,
\& Holberg, J. B. 2003, \aj, 125, 348

\bibitem[Liebert et al.(2005)]{2005ApJS..156...47L} Liebert, J., Bergeron, 
P., \& Holberg, J.~B.\ 2005, \apjs, 156, 47 

\bibitem[Luyten(1949)]{1949ApJ...109..528L} Luyten, W.~J.\ 1949, \apj, 109, 
528 

\bibitem[Luyten(1979a)]{1979lccs.book.....L} Luyten, W.~J.\ 1979, LHS
Catalogue (2nd ed.; Minneapolis: Univ. of Minnesota Press)

\bibitem[Luyten(1979b)]{nltt} Luyten, W.~J.\ 1979, New
Luyten Catalogue of Stars with Proper Motions Larger than Two Tenths
of an Arcsecond (Minneapolis: Univ. of Minnesota Press)

\bibitem[McCook \& Sion(1999)]{1999ApJS..121....1M} McCook, G.~P., \& Sion, 
E.~M.\ 1999, \apjs, 121, 1 


\bibitem[Oppenheimer et al.(2001)]{2001Sci...292..698O} Oppenheimer,
B.~R., Hambly, N.~C., Digby, A.~P., Hodgkin, S.~T., \& Saumon, D.\
2001, Science, 292, 698

\bibitem[Press et al.(1992)]{pressetal92} Press, W. H., Teukolsky, S. A., 
Vetterling, W. T., \& Flannery, B. P. 1992, Numerical Recipes in FORTRAN, 
2nd edition (Cambridge: Cambridge University Press), 644

\bibitem[Provencal et al.(1998)]{1998ApJ...494..759P} Provencal, J.~L., 
Shipman, H.~L., Hog, E., \& Thejll, P.\ 1998, \apj, 494, 759 

\bibitem[Pokorny et al.(2004)]{2004A&A...421..763P} Pokorny, R.~S., Jones, 
H.~R.~A., Hambly, N.~C., \& Pinfield, D.~J.\ 2004, \aap, 421, 763 

\bibitem[Salim et al.(2004)]{2004ApJ...601.1075S} Salim, S., Rich, R.~M., 
Hansen, B.~M., Koopmans, L.~V.~E., Oppenheimer, B.~R., \& Blandford, R.~D.\ 
2004, \apj, 601, 1075

\bibitem[Scholz et al.(2002)]{2002ApJ...565..539S} Scholz, R.-D., Szokoly, 
G.~P., Andersen, M., Ibata, R., \& Irwin, M.~J.\ 2002, \apj, 565, 539 

\bibitem[Skrutskie et al.(2006)]{2006AJ....131.1163S} Skrutskie, M.~F., et 
al.\ 2006, \aj, 131, 1163

\bibitem[Smart et al.(2003)]{2003A&A...404..317S} Smart, R.~L., et al.\ 
2003, \aap, 404, 317 

\bibitem[Subasavage et al.(2005a)]{2005AJ....129..413S} Subasavage, J.~P., 
Henry, T.~J., Hambly, N.~C., Brown, M.~A., \& Jao, W.\ 2005, \aj, 129, 413 

\bibitem[Subasavage et al.(2005b)]{2005AJ....130.1658S} Subasavage, J.~P., 
Henry, T.~J., Hambly, N.~C., Brown, M.~A., Jao, W.-C., \& Finch, C.~T.\ 
2005, \aj, 130, 1658 

%
%
%
%
%
%
%
%
%
%
%
%
%
%
%
%
%
%
%
%
%
%
%
%
%
%
%
%
%
%
%
%
\end{thebibliography}
\end{document}